\documentclass[11pt]{article}

\usepackage{amsmath, amssymb}
\usepackage[margin=0.75in]{geometry}
\usepackage{algorithm, algorithmic}
\usepackage{graphicx}

\usepackage{multirow}

\DeclareGraphicsExtensions{.pdf,.png,.jpg}

\allowdisplaybreaks
\usepackage[round]{natbib}

\usepackage{subcaption}
\usepackage{siunitx}

\begin{document}

%

%

\title{A Deterministic Global Optimization Method for Variational Inference}
\date{}
\author{ Hachem Saddiki \thanks{University of Massachusetts Amherst. Email: saddiki@math.umass.edu} \and Andrew C. Trapp \thanks{Worcester Polytechnic Institute. Email: atrapp@wpi.edu} \and Patrick Flaherty \thanks{University of Massachusetts Amherst. Email: flaherty@math.umass.edu}}

\maketitle

\begin{abstract}
Variational inference methods for latent variable statistical models have gained popularity because they are relatively fast, can handle large data sets, and have deterministic convergence guarantees.
However, in practice it is unclear whether the fixed point identified by the variational inference algorithm is a local or a global optimum.
Here, we propose a method for constructing iterative optimization algorithms for variational inference problems that are guaranteed to converge to the $\epsilon$-global variational lower bound on the log-likelihood.
We derive inference algorithms for two variational approximations to a standard Bayesian Gaussian mixture model (BGMM).
We present a minimal data set for empirically testing convergence and show that a variational inference algorithm frequently converges to a local optimum while our algorithm always converges to the globally optimal variational lower bound.
We characterize the loss incurred by choosing a non-optimal variational approximation distribution suggesting that selection of the approximating variational distribution deserves as much attention as the selection of the original statistical model for a given data set.
\end{abstract}

\section{Introduction} 

Maximum likelihood estimation of latent (hidden) variable models is computationally challenging because one must integrate over the latent variables to compute the likelihood.
Often, the integral is high-dimensional and computationally intractable so one must resort to approximation methods such as variational expectation-maximization~\citep{Beal2003a}.

The variational expectation-maximization algorithm alternates between maximizing an evidence lower bound (ELBO) on the log-likelihood with respect to the model parameters and maximizing the lower bound with respect to the variational distribution parameters.
Variational expectation-maximization is popular because it is computationally efficient and performs deterministic coordinate ascent on the ELBO surface.
However, variational inference has some statistical and computational limitations.
First, the ELBO is often multimodal, so the deterministic algorithm may converge to a local rather than a global optimum depending on the initial parameter estimates.
Second, the variational distribution is often chosen for computational convenience rather than for accuracy with respect to the posterior distribution, so the lower bound may be far from tight.
The magnitude of the cost of using a poor approximation is typically not known.
Here, we develop a deterministic global optimization algorithm for variational expectation-maximization to address these the first issue and quantify the optimal ELBO for an example model to address the second issue.


\paragraph{Related work} 

It is well-known that expectation-maximization algorithms can converge to a local optimum rather than a global optimum due to local modes or saddle-points in evidence lower bound on the log-likelihood function.
In a discussion for the~\citet{Dempster1977a} paper on expectation-maximization, \citet{Murray1977a} described a situation he frequently observed where the EM algorithm converged to a fixed point, but not the global optimum of the log-likelihood.
Murray provided a simple data set that reproduced the convergence problems and
advocated a solution that continues to be the common practice today -- restart the algorithm at multiple initial points and hope that it finds a global optimum.
Later, ~\citet{wu1983} described regularity conditions that ensure the EM algorithm converges to a global optimum.
However, these regularity conditions are difficult to check in practice.
Typically, the practical solution remains the one advocated by Murray -- restart the algorithm at many initial values and retain the parameter estimates that give the maximum across restarts.

\paragraph{Contributions}

In this article, we present a deterministic algorithm for variational inference that converges to the $\epsilon$-global optimum of the evidence lower bound on the log-likelihood. 
Global optimization methods such as the one we present here are important to study for several reasons.
First, in applications where collecting data is expensive or time-consuming (e.g. medical diagnostics and robotic planning in complex environments) we often have large, but not massive, data sets and we need an inference algorithm that maximizes the use of the information present in the sample while also being computationally tractable.
Second, global optimization methods provide new avenues for developing computationally-efficient inference algorithms that have not been extensively studied.
Lastly, a guaranteed global optimum parameter estimate allows us to benchmark approximate variational inference methods and approach the analysis of the statistical properties of those methods.

The remainder of this article is organized as follows.
In this section, we present the general variational inference problem statement for hierarchical exponential family latent variable models.
We summarize the properties and algorithmic steps for the GOP algorithm in Section~\ref{sec:gop_inference}.
We derive the GOP inference algorithm for two approximations of the Bayesian Gaussian mixture model in Section~\ref{sec:bgmm_pm} and Section~\ref{sec:bgmm_ga}.
In those sections, we describe experimental results comparing the GOP algorithm to variational-EM and BARON.
In the latter section, we compare the quality of the optimum for two variational models.

\paragraph{Problem statement}
Given a statistical model, the observed-data log-likelihood function, or simply the log-likelihood, is $l_{\mathcal{D}}(\phi) = \log p(\mathcal{D} \mid \phi)$, where $\mathcal{D}$ is the observed data set and $\phi$ is the set of model parameters.
The maximum likelihood parameter estimate is $\hat{\phi} \triangleq \arg \max_\phi l_{\mathcal{D}}(\phi)$.
If the model, $p(\mathcal{D} \mid \phi)$, is in the exponential family and all of the random variables are observed, the maximum likelihood optimization problem is convex and there is a unique maximum.
However, if the model has unobserved random variables, we must integrate over those random variables and the log-likelihood may be multi-modal.
When the integral is computationally intractable, we must resort to approximation methods to compute the maximum likelihood parameter estimate.


In a latent variable exponential family model, we introduce the unobserved or
latent variable $z_i$ with the conditional distribution, $p(x_i \mid z_i,\eta)$, and marginal distribution, $p(z_i|\theta)$.
Our observed data set is i.i.d. $\mathcal{D} = \{(x_i), i = 1, \ldots, N\}$.
The complete data set is $\mathcal{D}_C = \{(x_i, z_i), i = 1, \ldots, N\}$.
Taking $\phi \triangleq \{\eta, \theta\}$, the log-likelihood of the model is
\begin{equation}
  	l_{\mathcal{D}}(\phi) = \sum_{i=1}^N \log \int_{z_i} p(x, z_i \mid \phi) dz_i.
\end{equation}

Unfortunately, the integral over $z_i$ in the log-likelihood function is often computationally intractable.
By introducing an averaging distribution, $q(z \mid \xi)$, we can form a lower bound on the observed-data log-likelihood using Jensen's inequality,
\begin{align}
	l_\mathcal{D} (\phi) & = \sum_{i=1}^N \log \int_{z_i} p(x_i, z_i \mid \phi) dz_i \nonumber \\
	& = \sum_{i=1}^N \log \int_{z_i} q(z_i \mid \xi) \frac{p(x_i, z_i \mid \phi)}{q(z_i \mid \xi)} dz_i \nonumber\\
	& \geq \sum_{i=1}^N E_q\left[ \log p(x_i, z_i |\phi) \right] - \sum_{i=1}^N E_q\left[ \log q(z_i \mid \xi) \right] \label{eq:var_elbo} \nonumber\\ 
	& \triangleq \mathcal{L}(\xi, \phi) \nonumber.
\end{align}
Then, we can obtain an approximate maximum likelihood estimate by maximizing this evidence lower bound (ELBO) over both the averaging distribution and the model parameters, $\max_{\phi, \xi} \mathcal{L}(\xi, \phi)$.
The first term in the ELBO is the expected complete-data log-likelihood and the second term is the entropy of the variational distribution.

In the expected complete log-likelihood, there is a coupling between the variational distribution parameters, $\xi$, and the model parameters, $\phi$.
This coupling makes optimizing the ELBO difficult in practice because the resulting optimization problem is nonconvex.

\section{GOP Variational Inference} \label{sec:gop_inference}

The general Global OPtimization (GOP) algorithm was first developed by~\citet{Floudas1990a}. 
They extended the decomposition ideas of  \citet{Benders1962a} and
\citet{Geoffrion1972a} in the context of solving nonconvex optimization problems to global optimality~\citep{Floudas1993a, Floudas2000a}.
While the framework has been successfully used for problems in process design,
control, and computational chemistry, it has not, to our knowledge, been applied
to variational statistical inference.

\textbf{[Add reference to linear regression here.]}

Here, we extend the GOP algorithm for variational inference in hierarchical exponential family models.
In Section~\ref{sec:gop_problem_statement}, we state the problem conditions necessary for the GOP algorithm.
Then, we briefly review the mathematical theory for the GOP algorithm in Section~\ref{subsec:gop_theory}.
Finally, in Section~\ref{subsec:gop_algorithm}, we outline the general GOP algorithmic steps. 
The review of the GOP algorithm in this section is not novel and necessarily brief.
A more complete presentation of the general algorithm can be found in \citet{Floudas2000a}.

\subsection{Problem Statement}\label{sec:gop_problem_statement}

GOP addresses biconvex optimization problems that can be formulated as

\begin{equation}\label{eqn:gop_problem}
	\begin{split}
	\min_{\alpha, \beta} &\quad f(\alpha, \beta) 
	\\
	\text{s.t.} &\quad g(\alpha, \beta) \leq 0
	\\
	&\quad h(\alpha, \beta) = 0
	\\
	&\quad \alpha \in A, \beta \in B,
	\end{split}
\end{equation}
where $A$ and $B$ are convex compact sets, and $g(\alpha, \beta)$ and $h(\alpha, \beta)$ are vectors of inequality and equality constraints respectively.
We require the following conditions on \eqref{eqn:gop_problem}:
\begin{enumerate}
	\item $f(\alpha, \beta)$ is convex in $\alpha$ for every fixed $\beta$, and convex in $\beta$ for every fixed $\alpha$;
	\item $g(\alpha, \beta)$ is convex in $\alpha$ for every fixed $\beta$, and convex in $\beta$ for every fixed $\alpha$;
	\item $h(\alpha, \beta)$ is affine in $\alpha$ for every fixed $\beta$, and affine in $\beta$ for every fixed $\alpha$;
	\item first-order constraints qualifications (e.g. Slater's conditions) are satisfied for every fixed $\beta$.
\end{enumerate}
If these conditions are satisfied, we have a biconvex optimization problem and we can use the GOP algorithm to find the $\epsilon$-global optimum.

\paragraph{Partitioning and transformation of decision variables}

Variational inference is typically formulated as an alternating coordinate ascent algorithm where the evidence lower bound is iteratively maximized with respect to the model parameters and the variational distribution parameters.
However, the objective function construed under that partitioning of the decision variables is not necessarily biconvex.
Instead, we consider all of the decision variables (model parameters and variational parameters) in the variational inference optimization problem jointly.
Then, we are free to partition variables as we choose to ensure the GOP problem conditions are satisfied.

One may be able to transform variables in the original variational optimization problem, adding equality constraints as necessary, to satisfy the GOP problem conditions.
For larger problems and more complex models, the partitioning of the decision variables may not be apparent. 
\citet{Hansen1993a} proposed an algorithm to bilinearize quadratic and polynomial function problems, rational polynomials, and problems involving hyperbolic functions.

\paragraph{Biconvexity for exponential family hierarchical latent variable models}


The variational EM problem is a biconvex optimization problem under certain conditions.
We cast the variational EM problem into a biconvex convex optimization form by partitioning the model parameters, $\phi$, and the variational parameters $\xi$ into $\alpha$ and $\beta$ such that the GOP conditions are satisfied.
All of the functions are analytical and differentiable.
For two choices of approximating distribution for a Bayesian Gaussian mixture
model, we find a suitable partition of the decision variables.
However, what conditions guaranetee such a partition for general exponential
family hierarchical models remains an open question.

\subsection{GOP Theory}\label{subsec:gop_theory}

\paragraph{Primal problem}
Because the original problem~\eqref{eqn:gop_problem} satisfies the biconvexity conditions, fixing $\beta$ (say to $\beta^t \in B$) results in a convex subproblem,

\begin{equation}\label{eqn:gop_primal_problem}
	\begin{split}
	\min_{\alpha} &\quad f(\alpha, \beta^t) 
	\\
	\text{s.t.} &\quad g(\alpha, \beta^t) \leq 0
	\\
	&\quad h(\alpha, \beta^t) = 0
	\\
	&\quad \alpha \in A.
	\end{split}
\end{equation}
called the \textit{primal problem}.
The restriction of $\beta$ to a fixed value is equivalent to solving the original problem  with additional constraints.
Therefore, the solution of the primal problem provides an \textit{upper bound} on the original problem~\eqref{eqn:gop_problem}.
Since the problem is convex, it can be solved with any convex optimization solver.

In variational inference, this step would typically be considered the E-step if $\alpha$ was the variational parameters and the M-step if $\alpha$ was the model parameters.
However, we have partitioned the decision variables to formulate the problem as a biconvex optimization problem, therefore the direct EM-step interpretation is lost.

\paragraph{Relaxed dual problem}

We can write the original optimization problem \eqref{eqn:gop_problem} as

\begin{equation*}
	\begin{split}
	\min_{\beta} \min_{\alpha} &\quad f(\alpha, \beta) 
	\\
	\text{s.t.} &\quad g(\alpha, \beta) \leq 0
	\\
	&\quad h(\alpha, \beta) = 0
	\\
	&\quad \alpha \in A, \beta \in B,
	\end{split}
\end{equation*}
in order to cast the problem as inner and outer optimization problems using a \textit{projection} of the problem onto the space of $\beta$ variables:

\begin{eqnarray}\label{eqn:gop_inner_outer_problem}
	\min_{\beta} & v(\beta) & \nonumber\\
	\text{s.t.} &\quad v(\beta) = & \min_{\alpha \in A} f(\alpha, \beta) \\
	& & \text{s.t.} \quad g(\alpha, \beta) \leq 0 \nonumber \\
	& & \quad \quad \ h(\alpha, \beta) = 0 \nonumber \\
	& \quad \quad \beta \in B \cap V & \nonumber \\
	&\text{where} & V = \{ \beta : h(\alpha, \beta) = 0,\ g(\alpha, \beta) \leq 0 \text{ for some } \alpha \in A \}. \nonumber 
\end{eqnarray}
For a given $\beta^t$, the inner minimization problem in \eqref{eqn:gop_inner_outer_problem} is simply the primal problem.
Therefore, the function $v(\beta)$ can be defined for a set of solutions of the primal problem~\eqref{eqn:gop_primal_problem} for different values of $\beta$.
However, since $v(\beta)$ is defined implicitly \eqref{eqn:gop_inner_outer_problem} can be as difficult to solve as the original problem~\eqref{eqn:gop_problem}.

Because the original problem satisfies Slater's conditions, by nonlinear duality theory, the optimal objective function value of the primal problem \eqref{eqn:gop_primal_problem} for any $\beta = \beta^t$ is identical to the solution of its corresponding dual problem~\citep{},
\begin{equation*}
	\left\{
	\begin{array}{ll}
		\min\limits_{\alpha \in A} & f(\alpha, \beta^t) \\
		\text{s.t.} & g(\alpha, \beta^t) \leq 0 \\
		& h(\alpha, \beta^t) = 0
	\end{array}
	\right\}
	\equiv
	\sup_{\mu \geq 0, \lambda} \inf_{\alpha \in A} \left\{ f(\alpha, \beta^t) + \mu^\top g(\alpha, \beta^t) + \lambda^\top h(\alpha, \beta^t) \right\} \quad \forall\  \beta^t \in B \cap V.
\end{equation*}
Substituting into the primal problem~\eqref{eqn:gop_primal_problem} gives 
\begin{equation*}
	v(\beta) = \sup_{\mu \geq 0, \lambda} \inf_{\alpha \in A} \left\{ f(\alpha, \beta) + \mu^\top g(\alpha, \beta) + \lambda^\top h(\alpha, \beta) \right\} \quad \forall\ \beta \in B \cap V.
\end{equation*}
Using the definition of the supremum we can form the bound
\begin{equation*}
	v(\beta) \geq \inf_{\substack{\alpha \in A \\ \mu \geq 0, \lambda}} \left\{ f(\alpha, \beta) + \mu^\top g(\alpha, \beta) + \lambda^\top h(\alpha, \beta) \right\} \quad \forall\ \beta \in B \cap V.
\end{equation*}
Then, we have an equivalent problem to \eqref{eqn:gop_inner_outer_problem},
\begin{eqnarray}\label{eqn:gop_inner_outer_dual_problem}
	\min_{\beta} & v(\beta) & \nonumber\\
	\text{s.t.} &\quad v(\beta) \geq & \min_{\substack{\alpha \in A \\ \mu \geq 0,\ \lambda}} \left\{ f(\alpha, \beta) + \mu^\top g(\alpha, \beta) + \lambda^\top h(\alpha, \beta)  \right\} \\
	& \quad \quad \beta \in & B \cap V \nonumber \\
	&\text{where}\  V = & \{ \beta : h(\alpha, \beta) = 0,\ g(\alpha, \beta) \leq 0 \text{ for some } \alpha \in A \}. \nonumber 
\end{eqnarray}
The last two conditions in the dual problem \eqref{eqn:gop_inner_outer_dual_problem} define an implicit set of constraints making the solution of the problem difficult.

Dropping the last two conditions gives a \textit{relaxed dual problem},
\begin{equation}\label{eqn:gop_relaxed_dual_problem}
	\begin{split}
		\min_{\beta \in B,\ v_B} & v_B \\
		\text{s.t.} \quad & v_B \geq \min_{\alpha \in A} L(\alpha, \beta, \lambda, \mu) \quad \mu \geq 0,\ \lambda,
	\end{split}
\end{equation}
where $v_B \in \mathbb{R}$ and the Lagrange function for the primal problem~\eqref{eqn:gop_primal_problem} is 
\begin{equation}
	L(\alpha, \beta, \lambda, \mu) = f(\alpha, \beta) + \mu^\top g(\alpha, \beta) + \lambda^\top h(\alpha, \beta).
\end{equation}
We call the minimization problem in \eqref{eqn:gop_relaxed_dual_problem} the \textit{inner relaxed dual problem},
\begin{equation}\label{eqn:gop_inner_relaxed_dual_problem}
	L^*(\alpha^*, \beta, \lambda, \mu) = \min_{\alpha \in A} L(\alpha, \beta, \lambda, \mu),	
\end{equation}
where $\alpha^*$ denotes the value of the $\alpha$ variables at the optimal solution of the inner relaxed dual problem.

In summary, the primal problem \eqref{eqn:gop_primal_problem} contains \textit{more} constraints than the original problem because $\beta$ is fixed and so provides an \textit{upper bound} on the original problem.
The relaxed dual problem \eqref{eqn:gop_relaxed_dual_problem} contains \textit{fewer} constraints than the original problem and so provides a \textit{lower bound} on the original problem.
Because the primal and relaxed dual problems are formed from an inner-outer optimization problem, they can be used to form an iterative alternating algorithm to determine the global solution of the original problem.
Making the iteration $t$ explicit, the primal problem is
\begin{equation}
	\begin{split}
	\min_{\alpha} &\quad f(\alpha, \beta^t) 
	\\
	\text{s.t.} &\quad g(\alpha, \beta^t) \leq 0
	\\
	&\quad h(\alpha, \beta^t) = 0
	\\
	&\quad \alpha \in A,
	\end{split}
\end{equation}
and the relaxed dual problem is
\begin{equation}
	\begin{split}
		\min_{\beta \in B,\ v_B} & v_B \\
		\text{s.t.} \quad & v_B \geq \min_{\alpha \in A} L(\alpha, \beta, \lambda^t, \mu^t),
	\end{split}
\end{equation}
where
\begin{equation}
	L(\alpha, \beta, \lambda^t, \mu^t) = f(\alpha, \beta) + \mu^{t^\top} g(\alpha, \beta) + \lambda^{t^\top} h(\alpha, \beta).
\end{equation}
The key here is that information is passed from the primal problem to the dual problem through the Lagrange multipliers, $\{\lambda^t, \mu^t\}$, and information is passed from the dual problem to the primal problem through the optimal dual variables, $\beta^t$.

\paragraph{Relaxed dual problem decomposition} 
The form of the relaxed dual problem \eqref{eqn:gop_relaxed_dual_problem} is difficult to solve because it contains an inner relaxed dual problem that is parametric in $\beta$.
Here, we decompose the relaxed dual problem into a set of independent relaxed dual subproblems that are combined to find the solution of the relaxed dual problem.
The algorithm iterates between solving the primal problem and the relaxed dual problem in a way that provides guaranteed convergence to the global optimum.

Recall, the inner relaxed dual problem \eqref{eqn:gop_inner_relaxed_dual_problem} is 
\begin{equation*}
	L^*(\alpha^*, \beta, \lambda^t, \mu^t) = \min_{\alpha \in A} L(\alpha, \beta, \lambda^t, \mu^t).
\end{equation*}
Because the inner relaxed dual problem is convex in $\alpha$, we can form a lower bound by taking the first-order Taylor series approximation about some fixed $\alpha^t$,
\begin{equation}
	L^*(\alpha^*, \beta, \lambda^t, \mu^t) 
	\geq 
	\min_{\alpha \in A} L(\alpha, \beta, \lambda^t, \mu^t)\big\vert^{\text{lin}}_{\alpha^t}
\end{equation}
where
\begin{equation}
	L(\alpha, \beta, \lambda^t, \mu^t) \big \vert^{\text{lin}}_{\alpha^t}
	\triangleq L(\alpha^t, \beta, \lambda^t, \mu^t) + \left[ \nabla_\alpha L(\alpha, \beta, \lambda^t, \mu^t) \big\vert_{\alpha^t} \right]^T (\alpha - \alpha^t).
\end{equation}
For notational convenience we define
\begin{equation*}
	g_j^t(\beta) \triangleq \frac{\partial L(\alpha, \beta, \lambda^t, \mu^t)}{\partial \alpha_j}  \bigg \vert_{\alpha^t},
\end{equation*}
the $j$-th component of the gradient of the original problem Lagrange function with respect to $\alpha$ evaluated at $\alpha^t$.
Note that we are using $g$ to denote the inequality constraints in the original problem; this different function has a superscript $t$ that indicates the iteration.
Then, we can write the lower bound on the inner relaxed dual problem as
\begin{equation}
	\begin{split}
	\min_{\alpha \in A} L(\alpha, \beta, \lambda^t, \mu^t)\big\vert^{\text{lin}}_{\alpha^t}
	& =
	\min_{\alpha \in A} \left[ L(\alpha^t, \beta, \lambda^t, \mu^t) + \sum_j g_j^t(\beta) (\alpha_j - \alpha_j^t)  \right]
	\\
	& = 
	L(\alpha^t, \beta, \lambda^t, \mu^t) + \min_{\alpha_j \in A} \left[ \sum_j g_j^t(\beta) (\alpha_j - \alpha_j^t) \right].
	\end{split}
\end{equation}
For any fixed $\beta = \beta^d$ the summation and minimization can be exchanged,
\begin{equation}
	\min_{\alpha \in A} L(\alpha, \beta^d, \lambda^t, \mu^t)\big\vert^{\text{lin}}_{\alpha^t}
	=
	L(\alpha^t, \beta^d, \lambda^t, \mu^t) +  \sum_j \min_{\alpha_j \in A} g_j^t(\beta^d) (\alpha_j - \alpha_j^t).
\end{equation}
Because the $j$-th component of the second term on the right hand side is \textit{linear} in $\alpha_j$, the minimum will be at a bound of $\alpha_j \in A$.

The specific nature of the bound (lower or upper) is determined by the sign of $g_j^t(\beta) = \nabla_{\alpha_j} L(\alpha, \beta, \lambda^t, \mu^t) \big \vert_{\alpha^t}$.
There are two cases:
\begin{equation*}
\begin{cases}
	\text{if}\quad g_j^t(\beta) \geq 0 & \text{then}\quad \min\limits_{\alpha_j} g_j^t(\beta) (\alpha_j - \alpha_j^t) \geq g_j^t(\beta) (\alpha_j^L - \alpha_j^t) \\ 
	\text{if}\quad g_j^t(\beta) \leq 0 & \text{then}\quad \min\limits_{\alpha_j} g_j^t(\beta) (\alpha_j - \alpha_j^t) \geq g_j^t(\beta) (\alpha_j^U - \alpha_j^t).
\end{cases}	
\end{equation*}
where $\alpha_j^L$ and $\alpha_j^U$ are the lower and upper bounds of $\alpha_j \in A$ respectively.
We can write the two cases compactly as
\begin{equation*}
\min_{\alpha_j} g_j^t(\beta) (\alpha_j - \alpha_j^t) \geq g_j^t(\beta) (\alpha_j^B - \alpha_j^t),
\end{equation*}
where
\begin{equation*}
\alpha_j^B
=
\begin{cases}
	\alpha_j^L & \forall \beta : g_j^t(\beta) \geq 0 \\
	\alpha_j^U & \forall \beta : g_j^t(\beta) \leq 0.
\end{cases}	
\end{equation*}
So, for any fixed $\beta^d$, there exists a combination of bounds $B^*$ for the $\alpha$ variables such that
\begin{equation}\label{eqn:dual_bound}
\begin{split}
	\min_\alpha L(\alpha, \beta^d, \lambda^t, \mu^t) & \geq L(\alpha^t, \beta^d, \lambda^t, \mu^t) + \sum_j g_j^t(\beta^d) (\alpha_j^{B^*} - \alpha_j^t)
	\\
	& \geq L(\alpha^{B^*}, \beta^d, \lambda^t, \mu^t) \big \vert^{\text{lin}}_{\alpha^t}.
\end{split}	
\end{equation} 
The vector $B$ is the same size as $\alpha$ and each element $j$ indicates whether $\alpha_j$ is set at its upper or lower bound.
The variable $\alpha_j^{B^*}$ is the value of $\alpha_j$ at the upper/lower bound such that \eqref{eqn:dual_bound} is valid.
For any discrete $\beta^d$, we can obtain a lower bound on the inner relaxed dual problem $L^*(\alpha^*, \beta^d, \lambda^t, \mu^t)$ by taking the minimum of the linearized Lagrange function over all combinations of bounds $B \in CB$.
Since this is true for every $\beta^d$, it must be true for all $\beta$.

Computing the linearized Lagrange over all combinations of bounds requires that we compute $2^{|\alpha|}$ linearized Lagrange functions, which can be computationally expensive.
The Karush-Kuhn-Tucker (KKT) optimality conditions for the primal problem require that
\begin{equation*}
	g_j^t(\beta^t) = \nabla_{\alpha_j} L(\alpha, \beta^t, \lambda^t, \mu^t) \big \vert_{\alpha^t} = 0, \quad \forall j.
\end{equation*}
So, if the partial derivative of the primal Lagrange function with respect to $\alpha_j$ is found not to be a function of $\beta$, we do not need to consider the bounds of $\alpha_j$.
We reduce the computational burden by a factor of $2$ for every such variable.

Every $\alpha_j$ for which $g_j^t(\beta)$ is a function of $\beta$ is called a \textit{connected variable}.
The set of all such connected variables is 
\begin{equation}
	I_c^t \triangleq \left\{ j : g_j^t(\beta)\ \text{is a function of}\ \beta \right\}.	
\end{equation}
Therefore, the solution of the inner relaxed dual problem \eqref{eqn:gop_inner_relaxed_dual_problem} with Lagrange function replaced by its linearization about $\alpha^t$ depends only on the connected $\alpha$ variables, and we can write the inner relaxed dual minimization problem as
\begin{equation}
	\begin{split}
		\min_\alpha L(\alpha, \beta, \lambda^t, \mu^t) 
		&\geq 
		\min_{B \in CB} L(\alpha^t, \beta, \lambda^t, \mu^t) + \sum_{j \in I_c^t} g_j^t(\beta) (\alpha^B - \alpha_j^t)
		\\
		&\geq 
		L(\alpha^{B^*}, \beta, \lambda^t, \mu^t) \big \vert^{\text{lin}}_{\alpha^t}.
	\end{split}
\end{equation}

Now, we can solve for a lower bound of the inner relaxed dual problem by setting the connected variables at their bounds and minimizing the linearized primal Lagrange function.
Setting the connected variables at their bounds is equivalent to selecting a half-space defined by the cut (hyper-plane) $g_j^t(\beta) = 0$ in the domain of $\beta$.
Therefore, we are solving for a lower bound of the linearized Lagrange function in a unique region in the domain of $\beta$ for each $B \in CB$.
Since we solve for such a lower bound for all combinations of the connected variables, we cover the entire domain of $\beta$.

Since $g_i^t(\beta)$ can enter as either $\leq 0$ or $\geq 0$, we require that $g_j^t(\beta)$ be linear in $\beta$ to ensure all of the regions defined by $B \in CB$ are convex.
However, in general $g_j(\beta)$ is convex.
When $g_j(\beta)$ is convex, but not linear, we can linearize the Lagrange function with respect to $\beta$ by taking the first-order Taylor series approximation with respect to $\beta$ at $\beta^t$.
This linearization provides a lower bound on the Lagrange function and the resulting $g_j^t(\beta)$ cuts are linear in $\beta$.

\paragraph{Relaxed dual problem as a sequence of convex problems} 

We have shown that we can approximate the inner relaxed dual problem by a minimization over a finite combination of settings of $\alpha$ at its bounds.
Here, we show that we can solve the relaxed dual problem as a sequence of convex subproblems.
Finally, we show we achieve $\epsilon$-global convergence by iteratively alternating between solving the primal problem and one step in the sequence of relaxed dual subproblems.
These optimality results are based on proofs by \citet{Geoffrion1972a}.

Suppose we have a sequence of relaxed dual subproblems such that $v_B^{*^T}$ is the optimal value of the $T$-th relaxed dual problem.
\begin{equation}\label{eqn:gop_relaxed_dual_problem_T}
\mu_B^{*^T}	= \left\{
\begin{array}{cl}
	\min\limits_{\mu_B, \beta \in B} & \mu_B \\
	\text{s.t.} & \mu_B \geq \min\limits_{\alpha \in A} L(\alpha, \beta, \lambda^t, \mu^t),\quad  t = 1, \ldots, T-1 \\
	& \mu_B \geq \min\limits_{\alpha \in A} L(\alpha, \beta, \lambda^T, \mu^T).
\end{array}
\right\}
\end{equation}
Recall that for any $T$, the relaxed dual problem contains fewer constraints than the original dual problem and is a valid under-estimator of the original problem.
Further, since the relaxed dual problem at iteration $T$ contains more constraints than the relaxed dual problem at iteration $t < T$, the under-estimator is nondecreasing in $T$.
Now, we use the approach for solving the inner relaxed dual problem by setting $\alpha$ at its bounds to derive an iterative algorithm for \eqref{eqn:gop_relaxed_dual_problem_T}.

\subparagraph{Iteration 1}
For $t=1$ we have
\begin{equation*}
	\min_{\alpha \in A} L(\alpha, \beta, \lambda^1, \mu^1) \geq \min_{B \in CB}
	\left\{
	\begin{array}{c}
		L(\alpha^B, \beta, \lambda^1, \mu^1) \big \vert^{\text{lin}}_{\alpha^1} \\
		g_j^1(\beta) \leq 0 \ \text{if} \ \alpha_j^B = \alpha_j^U \\
		g_j^1(\beta) \geq 0 \ \text{if} \ \alpha_j^B = \alpha_j^L
	\end{array}
	\right\}.
\end{equation*}
As this holds for all $\beta$, it holds for the minimum over $\beta$,
\begin{equation*}
	\min_{\beta \in B} \left\{
	\min_{\alpha \in A} L(\alpha, \beta, \lambda^1, \mu^1) 
	\right\}
	\geq 
	\min_{\beta \in B} \left\{
	\min_{B \in CB}
	\left\{
	\begin{array}{c}
		L(\alpha^B, \beta, \lambda^1, \mu^1) \big \vert^{\text{lin}}_{\alpha^1}\\
		g_j^1(\beta) \leq 0 \ \text{if} \ \alpha_j^B = \alpha_j^U \\
		g_j^1(\beta) \geq 0 \ \text{if} \ \alpha_j^B = \alpha_j^L
	\end{array}
	\right\}
	\right\}.
\end{equation*}
Because the right hand side is only a function of $\beta$, the minimization operators can be interchanged,
\begin{equation*}
	\min_{\beta \in B} \left\{
	\min_{\alpha \in A} L(\alpha, \beta, \lambda^1, \mu^1) 
	\right\}
	\geq 
	\min_{B \in CB}
	\left\{
	\min_{\beta \in B} 
	\left\{
	\begin{array}{c}
		L(\alpha^B, \beta, \lambda^1, \mu^1) \big \vert^{\text{lin}}_{\alpha^1} \\
		g_j^1(\beta) \leq 0 \ \text{if} \ \alpha_j^B = \alpha_j^U \\
		g_j^1(\beta) \geq 0 \ \text{if} \ \alpha_j^B = \alpha_j^L
	\end{array}
	\right\}
	\right\},
\end{equation*}
or equivalently
\begin{equation*}
	\left\{
	\begin{array}{ll}
		\min\limits_{\beta \in B, \mu_B} & v_B \\
		\text{s.t.} & v_B \geq \min\limits_{\alpha \in A} L(\alpha, \beta, \lambda^1, \mu^1) 
	\end{array}
	\right\}
	\geq 
	\min_{B \in CB}
	\left\{
	\begin{array}{ll}
		\min\limits_{\beta \in B, \mu_B} & v_B \\
		\text{s.t.} & \mu_B \geq L(\alpha^B, \beta, \lambda^1, \mu^1) \big \vert^{\text{lin}}_{\alpha^1}\\
		& g_j^1(\beta) \leq 0 \ \text{if} \ \alpha_j^B = \alpha_j^U \\
		& g_j^1(\beta) \geq 0 \ \text{if} \ \alpha_j^B = \alpha_j^L
	\end{array}
	\right\}.
\end{equation*}
So, for iteration 1, we have an optimization problem that provides a valid lower bound on the inner relaxed dual problem.
For each setting of $\alpha$ at its bounds $B \in CB$, we solve a convex optimization problem -- the \textit{relaxed dual subproblem}.

\subparagraph{Iteration T}
Recall that $g_j^t(\beta)$ defines a cut in the domain of $\beta$ and all $j \in I_C^t$ cuts intersect at $\beta^t$ such that any point $\beta^d \in B$ lies in one and only one region defined by the cuts (with the exception of boundary points).
Further, recall that there is a one-to-one correspondence between $\alpha^B$, the primal problem variables set at their boundaries, and the region defined by the cuts, $g^t(\beta)$.
Now, let $UL(t, T)$ be the set of Lagrange functions and associated qualifying constraints from the $t$-th iteration whose qualifying constraints are satisfied at $\beta^T$, the current value of the relaxed dual problem decision variables.
The set of active qualifying constraints uniquely identifies $\alpha^{B_t}$, the primal problem variable bounds, where we denote the combination of bounds that uniquely identify the active qualifying constraints from iteration $t$ as $B_t$.

We can write the inner relaxed dual problems from \eqref{eqn:gop_relaxed_dual_problem_T} for iteration $t < T$ as
\begin{equation*}
	\left\{ 
	v_B \geq \min_{\alpha \in A} L(\alpha, \beta, \lambda^t, \mu^t)
	\right\}
	\equiv
	\left\{
	\begin{array}{cc}
		v_B \geq & L(\alpha^{B_t}, \beta, \lambda^t, \mu^t) \big \vert^{\text{lin}}_{\alpha^t} \\
		& g_j^t(\beta) \leq 0\ \text{if}\ \alpha_j^{B_t} = \alpha_j^U \\
		& g_j^t(\beta) \geq 0\ \text{if}\ \alpha_j^{B_t} = \alpha_j^L
	\end{array}
	\right\}
\end{equation*}
where $B_t$ is identified by $UL(t, T)$.
Since the $\alpha$ variables are set for all iterations $t < T$, the $\min_{B \in CB}$ operator only applies to the $T$-th iteration and we have
\begin{equation}\label{eqn:gop_relaxed_dual_problem_seq}
v_B^{*^T}	= \left\{
\begin{array}{cl}
	\min\limits_{v_B, \beta \in B} & v_B \\
	\text{s.t.} & 
	\begin{array}{ll}
		v_B \geq & L(\alpha^{B_t}, \beta, \lambda^t, \mu^t) \big \vert^{\text{lin}}_{\alpha^t} \\
		& g_j^t(\beta) \leq 0\ \text{if}\ \alpha_j^{B_t} = \alpha_j^U \\
		& g_j^t(\beta) \geq 0\ \text{if}\ \alpha_j^{B_t} = \alpha_j^L
	\end{array}
	,\quad  t = 1, \ldots, T-1 
	\\
	& v_B \geq \min\limits_{B \in CB} 
	\left\{
	\begin{array}{l}
		L(\alpha^B, \beta, \lambda^T, \mu^T) \big \vert^{\text{lin}}_{\alpha^T} \\
		g_j^T(\beta) \leq 0 \ \text{if} \ \alpha_j^B = \alpha_j^U \\
		g_j^T(\beta) \geq 0 \ \text{if} \ \alpha_j^B = \alpha_j^L
	\end{array}
	\right.
\end{array}
\right\}.
\end{equation}

Therefore, the relaxed dual problem can be solved as a series of convex problems.
Each iteration adds constraints to the relaxed dual problem ensuring that the lower bound is non-decreasing.
Since the relaxed dual problem is a valid lower bound on the global optimum for every iteration, it is a valid lower bound for the $T$-th iteration.
The proof that the upper bound provided by the primal and the lower bound provided by the relaxed dual converge can be found in \citep{Floudas2000a}[Theorem 3.6.1].

\subsection{GOP Algorithm Steps} \label{subsec:gop_algorithm}

\paragraph{Step 0 -- Initialize parameters.}
Initialize the upper bound $P^{UBD} = \infty$ and the lower bound $R^{LBD} = -\infty$.
Define lower and upper bounds for the primal decision variables $\alpha_j^L$ and $\alpha_j^U$ respectively.
Select a feasible initial value of the dual decision variables $\beta^1$.
Set the iteration counter to $T=1$.
Finally, select a convergence tolerance parameter $\epsilon$.

\paragraph{Step 1 -- Solve primal problem}
Solve the primal problem with the relaxed dual variables fixed at $\beta^T$ and store the optimal Lagrange multipliers $\lambda^T$ and $\mu^T$.
Update the upper bound $P^{UBD} = \min(P^{UBD}, P^T(\beta^T))$, where $P^T(\beta^T)$ is the solution of the primal problem.

\paragraph{Step 2 -- Select Lagrange functions from previous iterations}
While $T>1$, for $t = 1, \ldots, T-1$ identify the settings of the primal problem variables at their bounds $B_t$ such that all of the qualifying constraints from iteration $t$ are active at the current value of the relaxed dual variables $\beta^T$.
Select those qualifying constraints and their corresponding Lagrange function to be in the set of constraints $UL(t, T)$ of the current relaxed dual problem.

\paragraph{Step 3 -- Solve relaxed dual problem}
Determine the set of connected variables at iteration $T$, $I_C^T$.
For each setting of the connected variables at their bounds $\alpha^B$ where $B \in CB$, solve the \textit{relaxed dual problem}:
\begin{equation}
	\begin{split}
	\min_{v_B, \beta \in B} &\quad v_B  \\
	\text{s.t.} 
	& 
	\begin{array}{ll}
		\begin{array}{ll}
		v_B \geq & L(\alpha^{B_t}, \beta, \lambda^t, \mu^t) \big \vert^{\text{lin}}_{\alpha^t} \\
		& g_j^t(\beta) \leq 0\ \text{if}\ \alpha_j^{B_t} = \alpha_j^U \\
		& g_j^t(\beta) \geq 0\ \text{if}\ \alpha_j^{B_t} = \alpha_j^L
		\end{array}
	\end{array}
	\\
	& 
	\begin{array}{ll}
		\begin{array}{ll}
		v_B \geq & L(\alpha^{B}, \beta, \lambda^T, \mu^T) \big \vert^{\text{lin}}_{\alpha^T} \\
		& g_j^T(\beta) \leq 0\ \text{if}\ \alpha_j^{B} = \alpha_j^U \\
		& g_j^T(\beta) \geq 0\ \text{if}\ \alpha_j^{B} = \alpha_j^L,
		\end{array}
	\end{array}
	\end{split}
\end{equation}
where  $t = 1, \ldots, T-1$.

If feasible, store the solution in $v_B^{\text{stor}}(T, B)$ and $\beta^{\text{stor}}(T, B)$.
If not feasible, then store a null value.

\paragraph{Step 4 -- Update lower bound and $\beta^{(T+1)}$}
Select the minimum from the entire set $v_B^{\text{stor}}$, $v_B^{\text{min}}$.
Note that $v_B^{\text{stor}}$ contains the $2^{|I_C^T|}$ solutions of the relaxed dual problem from the current iteration and the solutions from all of the previous iterations.
Update the value of the lower bound $R^{LBD} \leftarrow v_B^{\text{min}}$.
Update the value of $\beta$ with the corresponding optimal relaxed dual decision variable $\beta^{(T+1)} \leftarrow \beta^{\text{min}}$.
Remove $v_B^{\text{min}}$ and $\beta^{\text{min}}$ from the stored sets so that they are not selected again in a subsequent iteration.

\paragraph{Step 5 -- Check for convergence}
Check if $R^{LBD} > P^{UBD} - \epsilon$. If the convergence criterion is satisfied, return the optimal parameter values, else increment the iteration counter $T \leftarrow T+1$ and go to step 1.

\section{BGMM -- Point Mass Approximation} \label{sec:bgmm_pm}

Here, we consider a variant of the Gaussian mixture model where cluster mean is endowed with a Gaussian prior.
We consider the variational model for the posterior distribution to be a point mass function for discrete latent random variables and a Dirac delta function for continuous random variables.
Variational expectation-maximization with a point-mass approximating distribution is exactly classical expectation-maximization \citep[p 337]{Gelman2013a}.

\subsection{GOP Algorithm Derivation} \label{bgmm_model}

\paragraph{Model structure}
A Gaussian mixture model with a prior distribution on the cluster means is 
\begin{equation}
\begin{split}
  M_k|\Gamma &\sim \text{Gaussian}(0, \Gamma)\ \text{for}\ k = 1, \ldots, K,\\
  Z_i|\pi &\sim \text{Categorial}(K, \pi),\ \text{for}\ i = 1, \ldots, N, \\
  Y_i|Z_{i}, M &\sim \text{Gaussian}(m_{z_i}, 1),\ \text{for}\ i = 1, \ldots, N,
\end{split}
\end{equation}
\\
with model parameters $\phi \triangleq \{ \pi, \Gamma \} = \{\pi_1, \ldots, \pi_K, \Gamma \}$, unobserved/latent variables $\{Z, M_1, \dots, M_K \}$, and observed variable $Y=y$.

Our inferential aim is the joint posterior distribution $Z_i, M \mid Y_i, \hat{\phi}$, for $i = 1, \ldots, N$, where $\hat{\phi}$ is the maximum likelihood parameter estimate.
The log likelihood for the model is  $l_\mathcal{D} (\phi) \triangleq \log f(y | \phi) $.

\paragraph{Variational distribution}
We take the variational distribution to be the following fully factorized (mean-field)  distribution
\begin{equation}
	\label{eqn:var_dist}
	q(z, m) = \prod_{i=1}^N q(z_i \mid \tau_i) \prod_{k=1}^K q(m_k | \nu_k),
\end{equation}
where
\begin{align*}
M_k | \nu_k &\sim  
	\begin{cases}
		1\ \text{if}\ m_k \equiv \nu_k \\
		0\ \text{otherwise}
	\end{cases}
\\
Z_i | \tau_i &\sim \text{Categorical}(K, \tau_i).
\end{align*}
We define $\xi \triangleq \{ \tau, \nu \}$ be the set of variational distribution parameters.

\paragraph{Variational lower bound for the log-likelihood}
Given the variational distribution \eqref{eqn:var_dist}, the evidence lower bound (ELBO) is found by applying Jensen's inequality.
The final ELBO is construed as a function of the model parameters, $\phi = \{\pi, \eta_m\}$, and the variational parameters $\xi = \{\tau, \nu\}$, where we have replaced $\Gamma$ by its natural parameter $\eta_m = - 1/(2\Gamma)$ to make the ELBO linear in the parameter.
The ELBO can be expanded to (see Appendix~\ref{sec:pmbgmm_elbo_derivation} for the full derivation) 
\begin{equation}\label{eqn:elbo}
\begin{split}
	\mathcal{L}(\phi, \xi) \propto
		- \frac{1}{2} \sum_{i=1}^N \sum_{k=1}^K \tau_{ik} ( y_i - \nu_k )^2
		+ \sum_{i=1}^N \sum_{k=1}^K \tau_{ik} \log \pi_k \\
	        + \eta_m \sum_{k=1}^K \nu_k^2
		+ \frac{K}{2} \log \left( -2 \eta_m \right)
		- \sum_{i=1}^N \sum_{k=1}^K \tau_{ik} \log \tau_{ik}.
\end{split}
\end{equation}

\paragraph{Partitioning and transforming decision variables}

The variational inference problem in standard form is a minimization problem, where the objective function is the negative ELBO. 

%
The objective function is not biconvex in the separation of variables into model parameters, $\phi = \{\pi, \eta_m\}$, and variational parameters, $\xi = \{\tau, \nu\}$ because of the first term involving $\tau_{ik}(y_i - \nu_k)^2$.
This term is cubic in the variational parameters and thus nonconvex.
However, if we separate $\tau$ and $\nu$, the first term is bi-convex. 
The second term is biconvex in $\pi$ and $\tau$.
The third term is biconvex in $\eta_m$ and $\nu$.
The remaining two terms are convex in $\eta_m$ and $\tau$ respectively.
So, we must separate $\tau$ from $\nu$, $\tau$ from $\pi$, and $\eta_m$ from $\nu$.
A partition that satisfies these conditions is $\alpha \triangleq \{ \nu, \pi \}$ and $\beta \triangleq \{ \tau, \eta_m \}$.
A proof that all four GOP algorithm conditions are satisfied is provided in Appendix~\ref{sec:gmm_biconvex_proof}.

Our selection of which variable to optimize over in the primal problem affects computational scalability.
If we choose to optimize over $\beta$ in the primal problem, we must solve at most $2^{|\beta|} = 2^{NK+1}$ relaxed dual subproblems for each iteration where each subproblem has $|\alpha| = 2K$ decision variables.
This is disadvantageous because the number of relaxed dual subproblems grows exponentially in the sample size.
Instead, we will optimize over $\alpha$ in the primal problem, which means we only need to solve at most $2^{|\alpha|} = 2^{2K}$ relaxed dual subproblems where each subproblem has $|\beta| = NK+1$ decision variables.
While each subproblem may be large, it will be a linear program which can be solved efficiently for millions of decision variables.

The variational inference problem is now 
\begin{equation}
\begin{split}
	\min_{\alpha, \beta} &\quad
		\frac{1}{2} \sum_{i=1}^N \sum_{k=1}^K \tau_{ik} ( y_i - \nu_k )^2
		- \sum_{i=1}^N \sum_{k=1}^K \tau_{ik} \log \pi_k \\
		&\quad - \frac{K}{2} \log \left( - 2 \eta_m \right)
		- \eta_{m} \sum_{k=1}^K \nu_k^2
		+ \sum_{i=1}^N \sum_{k=1}^K \tau_{ik} \log \tau_{ik}
	\\
	\text{s.t.} &\quad \sum_{k=1}^K \tau_{ik} - 1 = 0,\ \text{for}\ i=1,\ldots,N
	\\
	&\quad \sum_{k=1}^K \pi_k - 1 = 0,\quad \eta_m \leq 0, 
\end{split}
\end{equation}
where $\alpha \triangleq \{ \nu, \pi \}$ and $\beta \triangleq \{ \tau, \eta_m \}$.

\subsection{Experimental Results}
First, we examine the convergence of the upper and lower bounds on the ELBO for a small data set.
Then, we compare the accuracy and computational efficiency of GOP and variational inference algorithms.
Last, we compare the optimal solution of GOP and variational inference under many random initializations.

It is well-known that the expectation-maximization algorithm is only guaranteed
to converge to a fixed-point which may be arbitrarily far from a global optimum.
Others have offered anecdotes and evidence that it is actually rather common for expectation-maximization to converge to a local optimum or saddle point rather than a global optimum.
\citet{Murray1977a} described a simple data set that exhibits this local convergence issue for a Gaussian model.
We sought to identify a minimal data set for the (Bayesian) Gaussian mixture model that exhibits the local convergence problem.
\citet{archambeau2003} found that repeated observations and outliers are particularly problematic for expectation-maximization and the Gaussian mixture model. We consider these properties to construct the data set $y = [ -10, -10, 5, 25 ]$. 

We find that this minimal data set is useful for stress-testing any inference algorithm for the (Bayesian) Gassian mixture model.
%

\paragraph{Convergence experiment}

We examined the optimum ELBO values for GOP and variational EM (VEM) by randomly drawing initial model and variational parameters and running each algorithm to convergence.
The initial $\pi$ and $\tau_i$ are generated from a Dirichlet distribution with $\alpha = [1,1]$.
The initial $\Gamma$ is generated from a Gamma with shape parameter equal to the range of $y$.
The initial $\nu_k$ is drawn from uniform distribution over the range of $y$.
We drew 100 random initial values and we observed that VEM converges to a local optimum 87 times out of 100 restarts, while GOP converges to the global optimum for all 100 restarts. 
%

We verified the global optimal value using BARON, an independent global optimization algorithm~\citep{sahinidis1996baron,Tawarmalani2005a}.
These results show that GOP, empirically, always converges to the global optimum.
The VEM algorithm failed to converge to the global optimum more often than it succeeded.
These results quantify and agree with anecdotal evidence in the literature~\citep{Murray1977a}.

\paragraph{Computational time experiment}

We explore two aspects of computational efficiency: ``What is the behavior of the upper and lower bounds from the GOP algorithm across iterations?'' and ``How does the timing of the GOP algorithm compare to other local and global inference methods?''.

\begin{figure*}[h]
  \centering
  \begin{subfigure}[t]{0.45\textwidth}
    \centering
    \includegraphics[width=\textwidth]{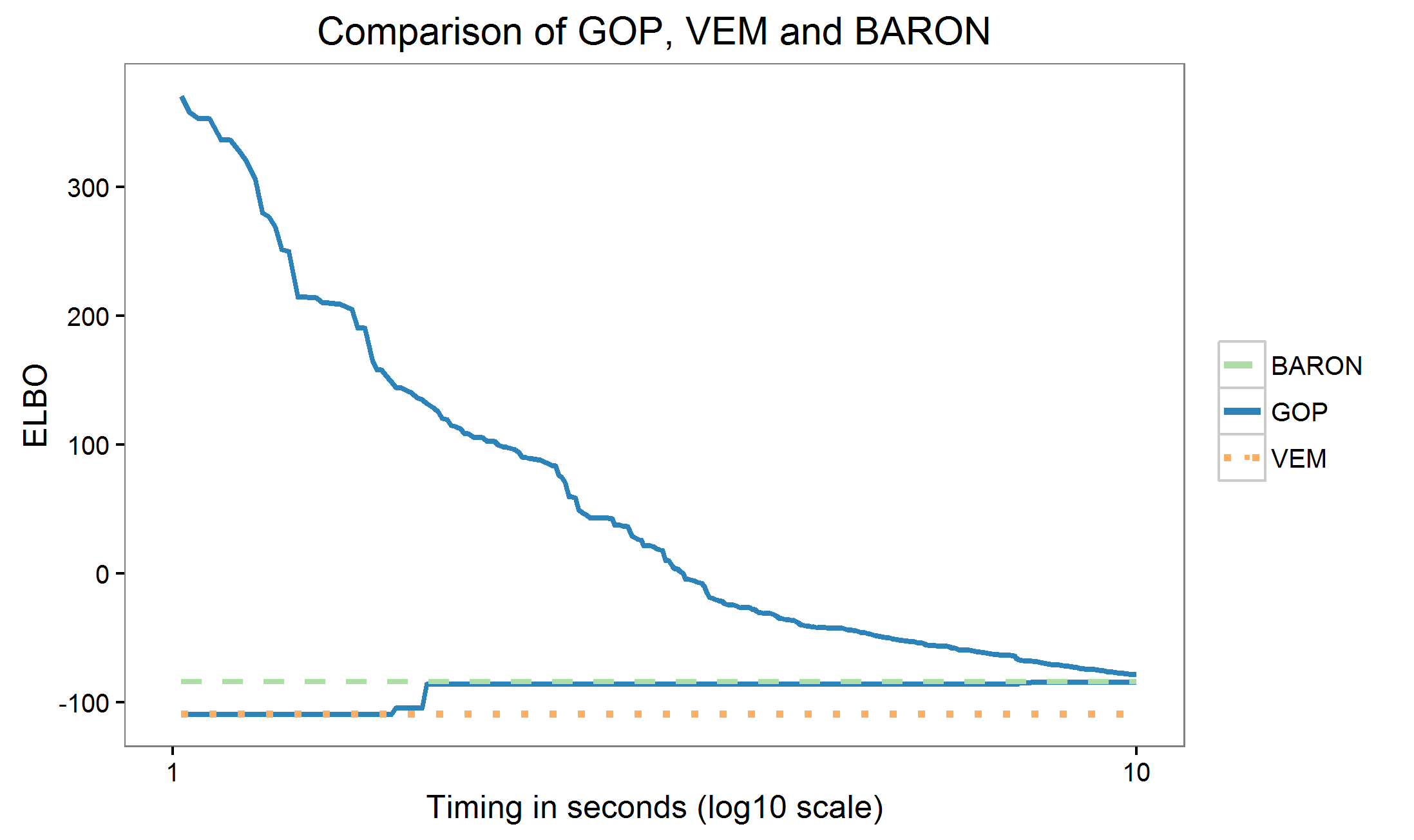}
    \caption{GOP iterations for a parameter initialization where VEM converges to a \textit{local} optimum.}
  \end{subfigure}
 \hfill
 \begin{subfigure}[t]{0.45\textwidth}
    \centering
    \includegraphics[width=\textwidth]{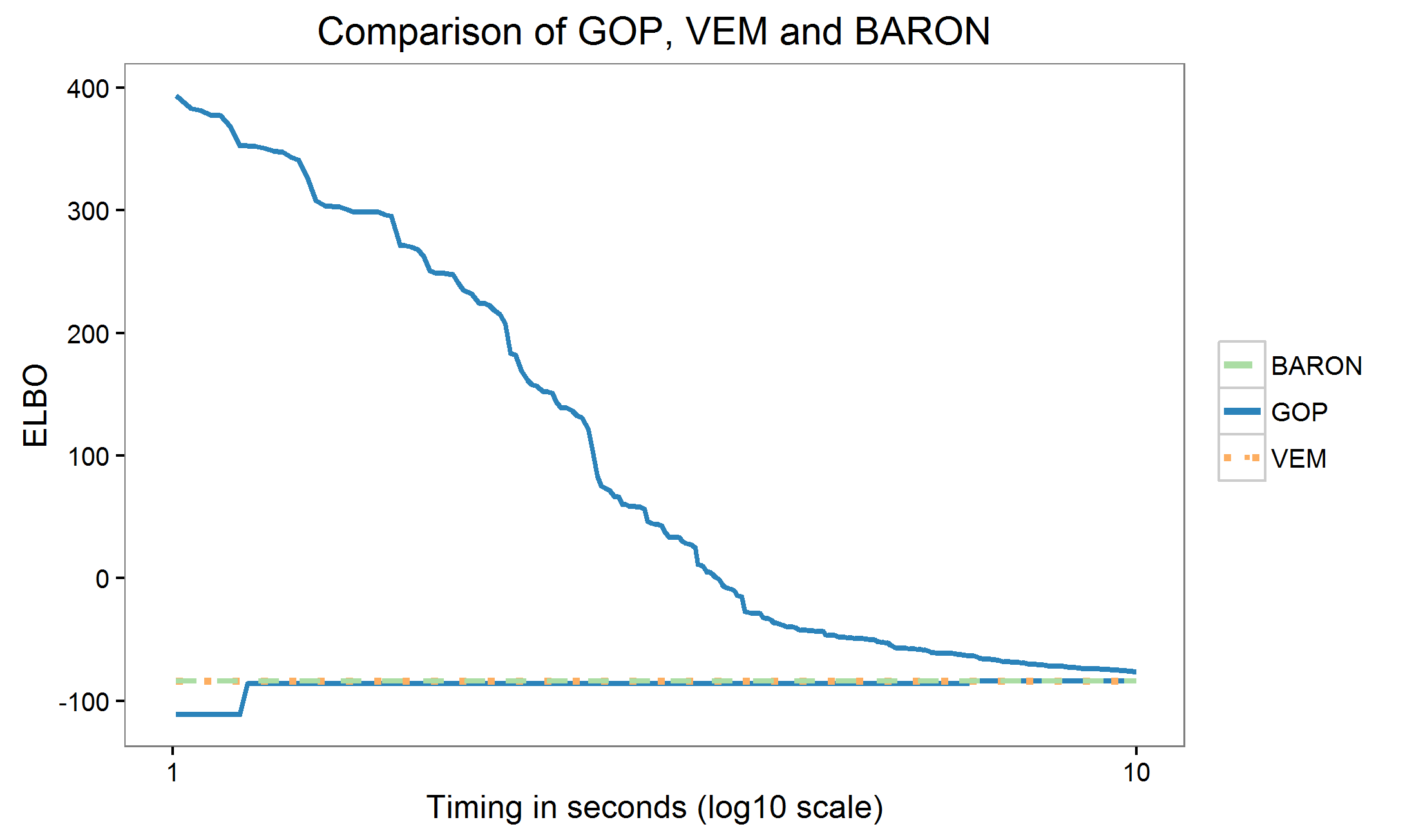}
    \caption{GOP iterations for a parameter initialization where VEM converges to a \textit{global} optimum.}
  \end{subfigure}
  \caption{Comparison of GOP, VEM, and BARON for two random initializations.}
  \label{fig:seed_comparison}
\end{figure*}

Figure~\ref{fig:seed_comparison} shows the convergence iterations for the GOP algorithm for two initial seeds.
For one of the seeds the VEM algorithm converges to a suboptimal local optimum Figure~\ref{fig:seed_comparison}(a) and for the other seed the VEM algorithm converges to a suboptimal local optimum Figure~\ref{fig:seed_comparison}(b).
The optimality of the solution is verified by BARON.
The globally optimal ELBO is $-84.04$ and the locally optimal ELBO is $-108.8$.

In early iterations, the GOP bounds include the local optimum, but at approximately 3 sec, the bound interval no longer covers the local optimum.
So, even though the GOP algorithm is slower than other local methods, one way it can be used is as a running oracle on an algorithm with only local guarantees.
When the GOP bound interval no longer covers the VEM solution, we can restart VEM until the solution found by VEM is covered by the GOP interval.
\begin{table}[H]
  \caption{Timing Comparison of GOP, Variational EM, and BARON Algorithms.}
  \label{tbl:compare_time}
  \centering
  \begin{tabular}{c c | c}
    \textbf{Algorithm} & & \textbf{Time} (s)\\
    \hline
    \textbf{GOP} &  & \\
      & $\epsilon = 1.0$ &  6.93 \\
      & $\epsilon = 0.1$ & 9.14 \\
      & $\epsilon = 0.01$ & 10.77 \\
    \textbf{BARON} & &   \\
      & $\epsilon = 1.0$ & 35\\
      & $\epsilon = 0.1$ & 38\\
      & $\epsilon = 0.01$ & 49\\
    \hline
    \textbf{Variational EM} &  &   \\
    & $\epsilon = 1.0$ &  0.0053 \\
      & $\epsilon = 0.1$ &  0.0071 \\
      & $\epsilon = 0.01$ & 0.0073 \\
  \end{tabular}
\end{table}
We compared the time for convergence to VEM and a standard global optimization algorithm, BARON for our minimal data set. The results are summarized in Table~\ref{tbl:compare_time}.

Our implementation of the GOP algorithm is faster than the global optimization algorithm, BARON and slower than VEM.
Indeed, VEM is several orders of magnitude faster.
However, the key point of this work is to demonstrate a method to obtain a global optimum.
Our implementation of GOP can be substanially enhanced by (1) using the reduction method that shrinks the decision variable search space at each iteration~\citep{Visweswaran1996-mr}, (2) using mixed-integer linear programming to identify the optimal relaxed dual sub-problem at each iteration~\citep{Visweswaran1996-mr}, and (3) using multi-threading to compute each master problem.
The multi-threading option is most appealing because each relaxed dual problem is completely independent, thereby suggesting how to drastically reduce the most expensive computation step.

\section{BGMM -- Gaussian Approximation} \label{sec:bgmm_ga}
In this section we explore the quality of the approximation provided by two different variational models for the Bayesian Gaussian mixture model.
The original model structure is the same as in Section~\ref{sec:bgmm_pm}, so we start with the variational model structure here.

\subsection{GOP Algorithm Derivation}

\paragraph{Variational distribution}
We take the variational distribution to be the following fully factorized (mean-field)  distribution
\begin{equation}
	\label{eqn:bgmm_v2_var_dist}
	q(z, m) = \prod_{i=1}^N q(z_i \mid \tau_i) \prod_{k=1}^K q(m_k | \nu_k, \gamma_k),
\end{equation}
where
\begin{align*}
M_k | \nu_k, \gamma_k &\sim  \mathcal{N}(\nu_k, \gamma_k), \\
Z_i | \tau_i &\sim \text{Categorical}(\tau_i).
\end{align*}
We define $\xi \triangleq \{ \tau, \nu, \gamma \}$ to be the set of variational distribution parameters.

\paragraph{Variational lower bound for the log-likelihood}
Given the variational distribution \eqref{eqn:bgmm_v2_var_dist}, the evidence lower bound (ELBO) is found by applying Jensen's inequality. 
The final ELBO is construed as a function of the model parameters, $\phi = \{\pi, \eta_m\}$, and the variational parameters $\xi = \{\tau, \nu, \gamma\}$, where we have replaced $\Gamma$ by its natural parameter $\eta_m = - 1/(2\Gamma)$ to make the ELBO linear in the parameter.
\begin{equation}
\label{eqn:gbgmm_gen_elbo}
	\begin{split}
		l_\mathcal{D} (\phi) \geq \mathcal{L}(\phi, \xi) &\triangleq E_q[\log p(y, z, m \mid \phi)] - E_q[\log q(z, m)] 
		\\
		&= \sum_{i=1}^N E_q[\log p(y_i \mid z_i, m)] + \sum_{i=1}^N E_q[\log p(z_i \mid \pi)] + \sum_{k=1}^K E_q[\log p(m_k \mid \Gamma)]
		\\
		&\quad - \sum_{i=1}^N E_q[\log q(z_i \mid \tau_i)] - \sum_{k=1}^K E_q[\log q(m_k \mid \nu_k, \gamma_k)].
	\end{split}
\end{equation}
The ELBO can be expanded (see Appendix~\ref{sec:gbgmm_elbo_derivation}) to
\begin{equation}\label{eqn:bgmm_v2_elbo}
  \begin{split}
    \mathcal{L}(\phi, \xi) & \propto
		- \frac{1}{2} \sum_{i=1}^N \sum_{k=1}^K \tau_{ik} ( y_i - \nu_k )^2
		- \frac{1}{2} \sum_{i=1}^N \sum_{k=1}^K \tau_{ik} \gamma_k
		+ \sum_{i=1}^N \sum_{k=1}^K \tau_{ik} \log \pi_k
		+ \frac{K}{2} \log \left( -2 \eta_m \right) 
		+ \eta_m \sum_{k=1}^K \nu_k^2\\
		&\quad + \eta_m \sum_{k=1}^K \gamma_k
		- \sum_{i=1}^N \sum_{k=1}^K \tau_{ik} \log \tau_{ik}
		+ \frac{1}{2} \sum_{k=1}^K \log (2 \pi e \gamma_k).
  \end{split}
\end{equation}

\paragraph{Partitioning and transforming decision variables}

The variational inference problem in standard form is a minimization problem, where the objective function is the negative ELBO. Using the same argument as in Section~\ref{sec:bgmm_pm}, we find a partition that satisfies the GOP conditions -- $\alpha \triangleq \{ \nu, \pi, \gamma \}$ and $\beta \triangleq \{ \tau, \eta_m \}$. A full proof that all four GOP algorithm conditions are satisfied is provided in the Appendix~\ref{sec:gbgmm_biconvex_proof}.


The variational inference problem is now 
\begin{align*}
	\min_{\alpha, \beta} &\quad
		\frac{1}{2} \sum_{i=1}^N \sum_{k=1}^K \tau_{ik} ( y_i - \nu_k )^2
		+ \frac{1}{2} \sum_{i=1}^N \sum_{k=1}^K \tau_{ik} \gamma_k \\
		&\quad - \sum_{i=1}^N \sum_{k=1}^K \tau_{ik} \log \pi_k
		- \frac{K}{2} \log \left( -2 \eta_m \right)
		- \eta_m \sum_{k=1}^K \nu_k^2  \\
		&\quad - \eta_m \sum_{k=1}^K \gamma_k
		+ \sum_{i=1}^N \sum_{k=1}^K \tau_{ik} \log \tau_{ik}
		- \frac{1}{2} \sum_{k=1}^K \log (2 \pi e \gamma_k)
	\\
	\text{s.t.} &\quad \sum_{k=1}^K \tau_{ik} - 1 = 0,\ \text{for}\ i=1,\ldots,N
	\\
	&\quad \sum_{k=1}^K \pi_k - 1 = 0,\quad \eta_m \leq 0, 
\end{align*}
where $\alpha \triangleq \{ \pi, \nu, \gamma \}$ and $\beta \triangleq \{ \tau, \eta_m \}$.

\subsection{Experimental Results}

In variational inference, the choice of approximating distribution is typically motivated by computational convenience rather than accuracy with respect to the posterior distribution.
In addition, the loss incurred by choosing a particular approximating distribution is generally not known.
The task of characterizing this loss is daunting because local methods such as VEM do not provide a performance guarantee.

\paragraph{Quality of optimal ELBO}
We investigate the quality of the optimal ELBO provided by (1) the point mass approximation detailed in Section~\ref{sec:bgmm_pm} and (2) the Gaussian approximation presented earlier in this section. 
We run GOP, adapted to both approximations, using the minimal data set described in~\ref{sec:bgmm_pm}. The globally optimal ELBO for point mass approximation and for Gaussian approximation is -84.04 and -82.75, respectively. 
We observe that the globally optimal ELBO for Gaussian approximation is higher than that of point mass approximation meaning that the Gaussian variational distribution is a better approximating distribution for our model compared to the point mass variational distribution.
This result shows that we do indeed incur a loss in accuracy when selecting an inferior approximating distribution suggesting that careful consideration must be made when choosing an approximating distribution.

\section{Conclusion}
There are several contributions of this article.
We describe an adaptation of a general-purpose biconvex global optimization method (GOP) for variational expectation-maximization problems.
We derive GOP inference algorithms for two variational approximations of the Bayesian Gaussian mixture model.
We provide a simple data set for which the variational evidence lower bound is multimodel that is useful for testing approximation methods.
We compare our GOP inference algorithm to one locally optimal algorithm (variational-EM) and one state-of-the-art global optimization algorithm (BARON) in terms of accuracy and computational time and show that GOP always converges to the global optimum and is not prohibitive in terms of computational time for the small data set tested.
We compare the globally optimal variational evidence lower bounds for two variational approximate distributions and show that the global variational evidence lower bound is better for a more flexible variational model.
The focus in this article is on the properties of the global optimization algorithm.
We describe several future improvements that can significantly improve the computational efficiency. 
\newpage
\appendix
\section{Appendix for GMM}

\subsection{Derivation of Log-likelihood Lower Bound}
\label{sec:gmm_elbo_derivation}

The variational lower bound is
\begin{equation}
  \begin{split}
    l_\mathcal{D} (\phi) \geq \mathcal{L}(\phi, \xi) 
	&\triangleq E_q[\log p(y, z \mid \phi)] - E_q[\log q(z \mid \xi)] 
	\\
	&= \sum_{i=1}^N E_q[\log p(y_i \mid z_i, \mu)] 
      + \sum_{i=1}^N E_q[\log p(z_i \mid \pi)] 
      - \sum_{i=1}^N E_q[\log q(z_i \mid \tau_i)].
  \end{split}
\end{equation}

Writing out each term of the bound gives
\begin{align*}
	E_q[\log p(y_i \mid z_i, \mu)] &= E_q \left[ -\frac{1}{2} \log (2 \pi) - \frac{1}{2} y_i^2 + \sum_{k=1}^K z_{ik} \mu_k - \frac{1}{2} \left( \sum_{k=1}^K z_{ik} \mu_k \right)^2 \right] \\
	&= -\frac{1}{2} \log (2 \pi) - \frac{1}{2} y_i^2 + \sum_{k=1}^K E_q \left[ z_{ik} \right] y_i - \frac{1}{2} \sum_{k=1}^K E_q \left[ z_{ik}^2 \right] \mu_k^2 
	\\
	E_q[\log p(z_i \mid \pi)] &= E_q \left[ \sum_{k=1}^K z_{ik} \log \pi_k \right] = \sum_{k=1}^{K} E_q[ z_{ik} ] \log \pi_k \\
	E_q[ \log q(z_i | \tau_i)] &= \sum_{k=1}^K \tau_{ik} \log \tau_{ik}.
\end{align*}
The ELBO requires expected values under the variational distribution for $E_q[ z_{ik} ]$ and $E_q[ z_{ik}^2 ]$.
We can simplify these expected values:
\begin{align*}
	E_q[ z_{ik} ] &= \tau_{ik} \\
	E_q[ z_{ik}^2 ] &= E_q\left[ \left( \sum_{k=1}^K z_{ik} \mu_k \right)^2 \right] = \sum_{k=1}^K E_q[z_{ik}^2] \mu_k^2 + 2 \sum_{k \neq k^\prime} E_q[z_{ik} z_{ik^\prime}] \mu_k \mu_{k^\prime} \\
	&= \sum_{k=1}^K \tau_{ik} \mu_k^2 + 2 \sum_{k \neq k^\prime} 0 \mu_k \mu_{k^\prime} = \sum_{k=1}^K \tau_{ik} \mu_k^2.
\end{align*}
Plugging these expectations back into the full ELBO gives us the result
\begin{equation}
	\mathcal{L}(\phi, \xi) \propto
		- \frac{1}{2} \sum_{i=1}^N \sum_{k=1}^K \tau_{ik} ( y_i - \mu_k )^2
		+ \sum_{i=1}^N \sum_{k=1}^K \tau_{ik} \log \pi_k
		- \sum_{i=1}^N \sum_{k=1}^K \tau_{ik} \log \tau_{ik}.
\end{equation}

\subsection{Variational Inference Algorithm}
\label{sec:gmm_variational_inference_algorithm}
The standard variational EM algorithm for the model is an alternating coordinate ascent algorithm on the ELBO.
The M-step maximizes the ELBO with respect to the model parameters, $\phi$.
The E-step maximizes the ELBO with respect to the variational distribution parameters, $\xi$.
Since the only variational quantity of interest is the expected value of the latent variables,$E_q[z_{ik}]$, variational inference is equivalent to the EM algorithm for this model.
Recall the ELBO is
\begin{equation}
  \mathcal{L}(\phi, \xi) = 
  -\frac{1}{2} \sum_{i=1}^N \sum_{k=1}^K \tau_{ik}  (y_i - \mu_k)^2 
  + \sum_{i=1}^N \sum_{k=1}^K \tau_{ik} \log \pi_k 
  - \sum_{i=1}^N \sum_{k=1}^K \tau_{ik} \log \tau_{ik},
\end{equation}
and the ELBO Lagrange function is 
\begin{equation}
  \begin{split}
  L(\alpha, \beta) &= 
    \frac{1}{2} \sum_{i=1}^N \sum_{k=1}^K \tau_{ik}  (y_i - \mu_k)^2 
    - \sum_{i=1}^N \sum_{k=1}^K \tau_{ik} \log \pi_k 
    + \sum_{i=1}^N \sum_{k=1}^K \tau_{ik} \log \tau_{ik}
    \\
    &\quad
    + \sum_{i=1}^N \lambda_{qi} \left( \sum_{k=1}^K \tau_{ik} - 1 \right)
    + \lambda_p \left( \sum_{k=1}^K \pi_k - 1 \right).
  \end{split}
\end{equation}

\paragraph{M-step}

We can maximize the ELBO with respect to $\pi$ by setting the derivative of the ELBO with respect to $\pi$ and solving for $\pi$.
Solving for the maximum of the resulting Lagrange function gives the update formula,
\begin{equation}
	\hat{\pi}_k = \frac{1}{N} \sum_{i=1}^N \tau_{ik},
\end{equation}
We see that the MLE for $\pi$ is the usual sample mean of $\tau$.

Setting the derivative of the ELBO Lagrange function with respect to $\mu_k$ and solving for $\mu_k$ gives the update formula,
\begin{equation}
	\hat{\mu}_k = \frac{\sum_{i=1}^N \tau_{ik} y_i}{\sum_{i=1}^N \tau_{ik}}.
\end{equation}

\paragraph{E-step}

Setting the derivative of the ELBO Lagrange function with respect to $\tau_{ik}$ and solving for $\tau_{ik}$ gives the update formula,
\begin{equation}
	\hat{\tau}_{ik} = \pi_k \exp\left\{ -\frac{1}{2} (y_i - \mu_k)^2 -1 -\lambda_{qi} \right\} = \frac{ \pi_k \exp\left\{ -\frac{1}{2} (y_i - \mu_k)^2\right\} } { \sum_{k=1}^K \pi_k \exp\left\{ -\frac{1}{2} (y_i - \mu_k)^2\right\} }.
\end{equation}
where we have solved for $\lambda_{qi}$ by plugging $\hat{\tau}_{ik}$ into its constraint.

\paragraph{Algorithm}
Putting the two update equations for the M-step together with the two update equations for the E-step gives the variational inference algorithm.

\begin{algorithm}
\caption{Gaussian mixture model variational inference algorithm}
\label{alg:gmm_variational}
\begin{algorithmic}
\STATE Initialize $\hat{\tau}$
\REPEAT 
  \STATE \COMMENT{M-step}
  \STATE Update $\hat{\pi}_k \leftarrow \frac{1}{N} \sum_{i=1}^N \hat{\tau}_{ik}$
  \STATE Update $\hat{\mu}_k = \frac{\sum_{i=1}^N \hat{\tau}_{ik} y_i}{\sum_{i=1}^N \hat{\tau}_{ik}}$
  \STATE \COMMENT{E-step}
  \STATE Update $\hat{\tau}_{ik} \frac{ \hat{\pi}_k \exp\left\{ -\frac{1}{2} (y_i - \hat{\mu}_k)^2\right\} } { \sum_{k=1}^K \hat{\pi}_k \exp\left\{ -\frac{1}{2} (y_i - \hat{\mu}_k)^2\right\} }$
\UNTIL{ELBO converges to fixed point}
\end{algorithmic}	
\end{algorithm}

\subsection{Proof that Gaussian Mixture Model ELBO satisfies GOP conditions}
\label{sec:gmm_biconvex_proof}

The objective function of interest is
\begin{equation}
  f(\phi, \xi) = 
  \frac{1}{2} \sum_{i=1}^N \sum_{k=1}^K \tau_{ik}  (y_i - \mu_k)^2 
  - \sum_{i=1}^N \sum_{k=1}^K \tau_{ik} \log \pi_k 
  + \sum_{i=1}^N \sum_{k=1}^K \tau_{ik} \log \tau_{ik},
\end{equation}
where $\phi = \{ \pi, \mu\}$ and $\xi=\{ \tau \}$.

The Hessian with respect to $\phi$ for fixed $\hat{\xi}$ is
\begin{equation}
	\nabla_{\phi}^2 f(\phi, \hat{\xi}) =
	\begin{bmatrix}
		\nabla^2_\pi f(\phi, \hat{\xi}) & 0 \\
		0 & \nabla^2_\mu f(\phi, \hat{\xi})	
	\end{bmatrix},
\end{equation}
where 
\begin{equation*}
  \nabla^2_\pi f(\phi, \hat{\xi}) = \text{diag} \left( \frac{1}{\pi_1} \sum_{i=1}^N \hat{\tau}_{i1}, \ldots, \frac{1}{\pi_K} \sum_{i=1}^N \hat{\tau}_{iK} \right)
\end{equation*}
and
\begin{equation*}
  \nabla^2_\mu f(\phi, \hat{\xi}) = \text{diag} \left( \sum_{i=1}^N \hat{\tau}_{i1}, \ldots, \sum_{i=1}^N \hat{\tau}_{iK} \right).
\end{equation*}
The matrix is diagonal and all of the diagonal elements are non-negative, so the Hessian is positive semidefinite and $f(\phi, \hat{\xi})$ is convex.

The Hessian with respect to $\xi$ for fixed $\hat{\phi}$ is
\begin{equation}
	\nabla_{\xi}^2 f(\hat{\phi}, \xi) =
	\nabla_{\tilde{\tau}}^2 f(\hat{\phi}, \tau) =
	\text{diag} \left( \frac{1}{\tau_{11}}, \ldots, \frac{1}{\tau_{NK}} \right),
\end{equation}
where $\tilde{\tau}$ is the matrix $\tau$ construed as a vector taken column-wise.
Again, the matrix is diagonal and all of the diagonal elements are non-negative, so the Hessian is positive semidefinite and $f(\hat{\phi}, \xi)$ is convex.

Therefore, the first condition of the GOP algorithm is satisfied.
The second condition is satisfied because we have no inequality constraints that are functions of both $\phi$ and $\xi$ -- all constraints can be absorbed into the sets $A,B$.
The third condition is satisfied because the equality constraints are each convex combinations and therefore affine.
The fourth condition (Slater's condition) is satisfied because a point exists
within the interior of the feasible region.
Interestingly, for the Gaussian mixture model, the problem is biconvex in the original separation of decision variables into model parameters and variational parameters.

\section{Appendix for BGMM -- Point Mass Approximation}

\subsection{Derivation of Log-Likelihood}
\label{sec:ll_derivation}

The likelihood function for the Bayesian Gaussian mixture model is
\begin{equation}
  p(y \mid \pi, \Gamma) = \prod_{i=1}^N p(y_i \mid \pi, \Gamma) = \prod_{i=1}^N \int_{z_i} \int_m p(y_i, z_i, m \mid \pi, \Gamma) dz_i dm.
\end{equation}

The log-likelihood function is then,
\begin{equation}
  l_y(\pi, \Gamma) = \sum_{i=1}^N \log \int_{z_i} \int_m p(y_i, z_i, m \mid \pi, \Gamma) dz_i dm.
\end{equation}
Since the model has a hierarchical structure, the joint density function can be expanded as a product of conditional distributions,
\begin{equation}
  \begin{split}
    l_y(\pi, \Gamma) &= \sum_{i=1}^N \log \int_{z_i} \int_m p(y_i \mid z_i, m) p(z_i \mid \pi) p(m \mid \Gamma) dz_i dm \\
    & = \sum_{i=1}^N \log \int_{z_i} \int_{m_1} \cdots \int_{m_K} p(y_i \mid z_i, m) p(z_i \mid \pi) \prod_{k=1}^K p(m_k \mid \Gamma) dz_i dm_1 \cdots dm_K.
  \end{split}
\end{equation}
The integral with respect to $z_i$ can be replaced with a summation since it is finite,
\begin{equation}
  l_y(\pi, \Gamma) = \sum_{i=1}^N \log \sum_{z_i=1}^K \int_{m_1} \cdots \int_{m_K} p(y_i \mid z_i, m) p(z_i \mid \pi) \prod_{k=1}^K p(m_k \mid \Gamma) dm_1 \cdots dm_K.
\end{equation}
The conditional density functions are:
\begin{eqnarray*}
  p(y_i \mid z_i, m) &=& \frac{1}{\sqrt{2 \pi}} \exp \left( -\frac{1}{2} (y_i - m_{z_i})^2 \right),\\
  p(z_i \mid \pi) &=& \pi_{z_i}, \text{and}\\
  p(m_k \mid \Gamma) &=& \frac{1}{\sqrt{2 \pi \Gamma}} \exp \left( -\frac{1}{2 \Gamma} m_k^2 \right).
\end{eqnarray*}
The conditional density functions can be substituted in to the log-likelihood to give
\begin{equation}
  l_y(\pi, \Gamma) = \sum_{i=1}^N \log \sum_{z_i=1}^K \int_{m_1} \cdots \int_{m_K}
  \frac{1}{\sqrt{2 \pi}} \exp \left( -\frac{1}{2} (y_i - m_{z_i})^2 \right)
  \pi_{z_i}
  \prod_{k=1}^K \frac{1}{\sqrt{2 \pi \Gamma}} \exp \left( -\frac{1}{2 \Gamma} m_k^2 \right) dm_1 \cdots dm_K.
\end{equation}
Now, if $z_i=k$ the integrals for $m_{k^\prime}$ where $k^\prime \ne k$ do not involve the density for $y_i$ or $z_i$ and are constant.
The kernels for those integrals are then $p(m_{k^\prime} \mid \Gamma)$ and the integrals are equal to one.
Therefore, the log-likelihood can be simplified as
\begin{equation}
  l_y(\pi, \Gamma) = \sum_{i=1}^N \log \sum_{z_i=1}^K g(z_i),
\end{equation}
where
\begin{equation}
  \begin{split}
    g(z_i) &= \frac{1}{2 \pi \sqrt{\Gamma}} \pi_{z_i} \int_{m_{z_i}} \exp \left\{ -\frac{1}{2} (y_i^2 -2m_{z_i}y_i + m_{z_i}^2 +\frac{m_{z_i}}{\Gamma}  \right\} dm_{z_i} \\
    & = \frac{1}{2 \pi \sqrt{\Gamma}} \pi_{z_i} \int_{m_{z_i}} \exp \left\{ -\frac{1}{2} \left[ \frac{\Gamma+1}{\Gamma} \right] \left( m_{z_i} - y_i \frac{\Gamma}{\Gamma+1} \right)^2 \right\} dm_{z_i} \cdot \exp \left\{ \frac{1}{2} \frac{\Gamma}{\Gamma+1} y_i^2 - \frac{1}{2} y_i^2 \right\} .
  \end{split}
\end{equation}
After completing the square, the integral can be substituted by the normalizing factor for a Gaussian distribution,
\begin{equation}
  g(z_i) = \frac{1}{2 \pi \sqrt{\Gamma}} \pi_{z_i} \sqrt{2 \pi \frac{\Gamma}{\Gamma+1}} \exp \left\{ -\frac{1}{2} y_i^2 \frac{1}{\Gamma+1} \right\}.
\end{equation}
Simplifying and substituting $g(z_i)$ back into the log-likelihood gives
\begin{equation}
  l_y(\pi, \Gamma) = \sum_{i=1}^N \log \sum_{z_i=1}^K \left[ \frac{1}{\sqrt{2 \pi (\Gamma + 1)}} \pi_{z_i} \exp \left\{ - \frac{1}{2 (\Gamma + 1)} y_i^2 \right\} \right].
\end{equation}
The only term left involving $z_i$ is $\pi_{z_i}$, and $\sum_{z_i=1}^K \pi_{z_i}=1$, so the parameter $pi$ drops out of the log-likelihood,
\begin{equation}
  l_y(\pi, \Gamma) = -\frac{N}{2}\log(2\pi) - \frac{N}{2} \log (\Gamma+1) - \frac{1}{2 (\Gamma+1)} \sum_{i=1}^N y_i^2 \propto -N \log(\Gamma+1) - \frac{1}{\Gamma+1} \sum_{i=1}^N y_i^2.
\end{equation}
Taking the derivative, setting it equal to zero and solving for $\Gamma$ gives the maximum likelihood estimate
\begin{equation}
  \hat{\Gamma} = \frac{1}{N} \sum_{i=1}^N y_i^2 - 1,
\end{equation}
and substituting back into the log-likelihood gives the value of the maximum log-likelihood.

\subsection{Derivation of Log-likelihood Lower Bound}
\label{sec:pmbgmm_elbo_derivation}

The variational lower bound is
\begin{equation}
	\begin{split}
		l_\mathcal{D} (\phi) \geq \mathcal{L}(\phi, \xi) &\triangleq E_q[\log p(y, z, m \mid \phi)] - E_q[\log q(z, m)] 
		\\
		&= \sum_{i=1}^N E_q[\log p(y_i \mid z_i, m)] + \sum_{i=1}^N E_q[\log p(z_i \mid \pi)] + \sum_{k=1}^K E_q[\log p(m_k \mid \Gamma)]
		\\
		&\quad - \sum_{i=1}^N E_q[\log q(z_i \mid \tau_i)] - \sum_{k=1}^K E_q[\log q(m_k \mid \nu_k)].
	\end{split}
\end{equation}
We can write out each term in the ELBO:
\begin{align*}
	E_q[\log p(y_i \mid z_i,m)] &= E_q \left[ -\frac{1}{2} \log (2 \pi) - \frac{1}{2} y_i^2 + y_i \sum_{k=1}^K z_{ik} m_k - \frac{1}{2} \left( \sum_{k=1}^K z_{ik} m_k \right)^2 \right] \\
	&= -\frac{1}{2} \log (2 \pi) - \frac{1}{2} y_i^2 + E_q \left[ \sum_{k=1}^K z_{ik} m_k \right] y_i - \frac{1}{2} E_q \left[ \left( \sum_{k=1}^K z_{ik} m_k \right)^2 \right]
	\\
	E_q[\log p(z_i \mid \tau_i)] &= E_q \left[ \sum_{k=1}^K z_{ik} \log \tau_{ik} \right]
	=  \sum_{k=1}^K E_q [ z_{ik} ] \log \tau_{ik} 
	\\
	E_q[\log p(m_k | \eta_m)] &= E_q \left[ -\frac{1}{2} \log (2 \pi) + \eta_m m_k^2 + \log ( - 2 \eta_m ) \right] \\
	&= -\frac{1}{2} \log (2 \pi) + \eta_m E_q \left[ m_k^2 \right] + \frac{1}{2} \log ( - 2 \eta_m ) 
	\\
	E_q[ \log q(z_i | \tau_i)] &= \sum_{k=1}^K \tau_{ik} \log \tau_{ik} 
	\\
	E_q[ \log q(m_k | \nu_k)] &= \int q(m_k | \nu_k) \log q(m_k | \nu_k) dm_k = \log q(\nu_k | \nu_k) = \log(1) = 0,	
\end{align*}
where we have substituted $\eta_m \triangleq -(2 \Gamma)^{-1}$.
The ELBO requires expected values under the variational distribution for $E_q \left[ \sum_{k=1}^K z_{ik} m_k \right]$, $E_q \left[ \left( \sum_{k=1}^K z_{ik} m_k \right)^2 \right]$, $E_q[z_{ik}]$, and $E_q[m_k^2]$.
We can simplify these expected values:
\begin{align*}
	E_q \left[ \sum_{k=1}^K z_{ik} m_k \right] &= \sum_{k=1}^K E_q[z_{ik}] E_q[m_k] = \sum_{k=1}^K \tau_{ik} \nu_k, \\
	E_q\left[ \left( \sum_{k=1}^K z_{ik} m_k\right)^2 \right] &= \sum_{k=1}^K E_q[z_{ik}^2] E_q[m_k^2] + 2 \sum_{k \neq k^\prime} E_q[z_{ik} z_{ik^\prime}] E_q[m_k m_{k^\prime}], \\
	&= \sum_{k=1}^K \tau_{ik} \nu_k^2 + 2 \sum_{k \neq k^\prime} 0 \nu_k \nu_{k^\prime} = \sum_{k=1}^K \tau_{ik} \nu_k^2, \\
	E_q[z_{ik}] &= \tau_{ik}, \\
	E_q[m_k^2] &= \nu_k^2.
\end{align*}
Substituting these expectations back into the full ELBO gives us the result
\begin{equation}
	\mathcal{L}(\phi, \xi) \propto
		- \frac{1}{2} \sum_{i=1}^N \sum_{k=1}^K \tau_{ik} ( y_i - \nu_k )^2
		+ \sum_{i=1}^N \sum_{k=1}^K \tau_{ik} \log \pi_k
		+ \eta_m \sum_{k=1}^K \nu_k^2
		+ \frac{K}{2} \log \left( -2\eta_m \right)
		- \sum_{i=1}^N \sum_{k=1}^K \tau_{ik} \log \tau_{ik}.
\end{equation}

\subsection{Variational Inference Algorithm}
\label{sec:pmbgmm_variational_inference_algorithm}

The standard variational EM algorithm for the model is an alternating coordinate ascent algorithm on the ELBO.
The M-step maximizes the ELBO with respect to the model parameters, $\phi$.
The E-step maximizes the ELBO with respect to the variational distribution parameters, $\xi$.

\paragraph{M-step}

We can maximize the ELBO with respect to $\pi$ by setting the derivative of the ELBO function with respect to $\pi$ and solving for $\pi$.
However, we also require that $\sum_{k=1}^K \pi_k = 1$, so we add a Lagrange multiplier to enforce this constraint.
Solving for the maximum of the resulting Lagrange function gives the update formula,
\begin{equation}
	\hat{\pi}_k = \frac{1}{N} \sum_{i=1}^N \tau_{ik},
\end{equation}
We see that the MLE for $\pi$ is the usual sample mean of $\tau$.

Setting the derivative of the ELBO Lagrange function with respect to $\eta_{m}$ and solving for $\eta_{m}$ gives the update formula,
\begin{equation}
	\hat{\eta}_m \triangleq - K \left( 2 \sum_{k=1}^K \nu_k^2 \right)^{-1}.
\end{equation}
This natural parameter estimate gives the usual MLE estimate for the corresponding moment parameter $\hat{\Gamma} = \frac{1}{K} \sum_{k=1}^K \nu_k^2$.

\paragraph{E-step}

Setting the derivative of the ELBO Lagrange function with respect to $\tau_{ik}$ and solving for $\tau_{ik}$ gives the update formula,
\begin{equation}
	\hat{\tau}_{ik} \triangleq \frac{1}{\lambda_{qi}} \exp \left( \nu_k - \frac{1}{2} \nu_k^2 + \eta_{z_k} \right) 
	= \frac{ \exp \left( \nu_k - \frac{1}{2} \nu_k^2 + \eta_{z_k} \right) }{ \sum_{k=1}^K \exp \left( \nu_k - \frac{1}{2} \nu_k^2 + \eta_{z_k} \right) }.
\end{equation}
where we have solved for $\lambda_{qi}$ by plugging $\hat{\tau}_{ik}$ into its constraint.
The update for $\tau_{ik}$ can be rewritten using the moment parameters as
\begin{equation*}
	\hat{\tau}_{ik} = \frac{ \pi_k \exp \left( -\frac{1}{2} (y_i - \nu_k)^2 \right) }{ \sum_{k=1}^K \pi_k \exp \left( -\frac{1}{2} (y_i - \nu_k)^2 \right) }.
\end{equation*}
This form shows that $\hat{\tau}_{ik}$ is proportional to the product of the prior for belonging to the $k$-th cluster and the likelihood of sampling $y_i$ from a Gaussian with mean $\nu_k$.

Setting the derivative of the ELBO Lagrange function with respect to $\nu_{k}$ and solving for $\nu_{k}$ gives the update formula,
\begin{equation}
	\hat{\nu}_{k} \triangleq \frac{ \sum_{i=1}^N y_i \tau_{ik} }{ \sum_{i=1}^N \tau_{ik} - 2 \eta_m }.
\end{equation}
The update for $\nu_k$ can be rewritten using the moment parameters as
\begin{equation*}
	\hat{\nu}_{k} = \frac{ \sum_{i=1}^N y_i \tau_{ik} }{ \sum_{i=1}^N \tau_{ik} + \Gamma^{-1} }.
\end{equation*}
This form shows that $\hat{\nu}_{k}$ is the sample mean of the weighted data assigned to the $k$-th cluster when the prior precision $\Gamma^{-1} = 0$.
This value of the precision corresponds to an improper prior on the cluster means and reduces the Bayesian Gaussian mixture model to a standard Gaussian mixture model.
When $\Gamma^{-1}$ is greater than zero $\hat{\nu}_k$ is regularized (shrunken) towards the prior mean, which in this model is zero.

\paragraph{Algorithm}
Putting the two update equations for the M-step together with the two update equations for the E-step gives the variational inference algorithm.

\begin{algorithm}
\caption{BGMM point mass variational approximate inference algorithm}
\label{alg:pmbgmm_variational}
\begin{algorithmic}
\STATE Initialize $\hat{\tau}$, $\hat{\nu}$
\REPEAT 
  \STATE \COMMENT{M-step}
  \STATE Update $\hat{\pi}_k \leftarrow \frac{1}{N} \sum_{i=1}^N \hat{\tau}_{ik}$
  \STATE Update $\hat{\eta}_m \leftarrow - K \left( 2 \sum_{k=1}^K \hat{\nu}_k^2 \right)^{-1}$
  \STATE \COMMENT{E-step}
  \REPEAT
    \STATE Update $\hat{\tau}_{ik} \leftarrow \frac{ \hat{\pi}_k \exp \left( -\frac{1}{2} (y_i - \hat{\nu}_k)^2 \right) }{ \sum_{k=1}^K \hat{\pi}_k \exp \left( -\frac{1}{2} (y_i - \hat{\nu}_k)^2 \right) }$
    \STATE Update $\hat{\nu}_{k} \leftarrow \frac{ \sum_{i=1}^N y_i \hat{\tau}_{ik} }{ \sum_{i=1}^N \hat{\tau}_{ik} - 2 \hat{\eta}_m }$
  \UNTIL{variational parameters converge}
\UNTIL{ELBO converges to fixed point}
\end{algorithmic}	
\end{algorithm}

\subsection{Proof of GOP conditions}
\label{sec:pmbgmm_biconvex_proof}

The objective function of interest is
\begin{equation}\label{eqn:pmbgmm_gop_objective}
	f(\alpha, \beta) = 
		\frac{1}{2} \sum_{i=1}^N \sum_{k=1}^K \tau_{ik} ( y_i - \nu_k )^2
		- \sum_{i=1}^N \sum_{k=1}^K \tau_{ik} \log \pi_k
		- \frac{K}{2} \log \left( - 2 \eta_m \right)
		- \eta_{m} \sum_{k=1}^K \nu_k^2
		+ \sum_{i=1}^N \sum_{k=1}^K \tau_{ik} \log \tau_{ik}
\end{equation}
where $\alpha \triangleq \{ \nu, \pi \}$ and $\beta \triangleq \{ \tau, \eta_m \}$.

The Hessian with respect to $\beta$ for fixed $\hat{\alpha}$ is
\begin{equation}
	\nabla_{\beta}^2 f(\hat{\alpha}, \beta) =
	\begin{bmatrix}
		\nabla^2_{\tilde{\tau}} f(\hat{\alpha}, \beta) & 0 \\
		0 & \nabla^2_{\eta_m} f(\hat{\alpha}, \beta)	
	\end{bmatrix},
\end{equation}
where 
\begin{equation*}
  \nabla^2_{\tilde{\tau}} f(\hat{\alpha}, \beta) = \text{diag} \left( \frac{1}{\tau_{11}}, \ldots, \frac{1}{\tau_{NK}} \right),
\end{equation*}
and
\begin{equation*}
  \nabla^2_{\eta_m} f(\hat{\alpha}, \beta) = \frac{K}{2 \eta_m^2}.
\end{equation*}
We define $\tilde{\tau}$ to be the matrix $\tau$ construed as a vector taken column-wise.
Since the matrix is diagonal and all of the diagonal elements are non-negative, the Hessian is positive semidefinite and $f(\hat{\alpha}, \beta)$ is convex.

The Hessian with respect to $\alpha$ for fixed $\hat{\beta}$
\begin{equation}
	\nabla_{\alpha}^2 f(\alpha, \hat{\beta}) =
	\begin{bmatrix}
		\nabla^2_{\pi} f(\alpha, \hat{\beta}) & 0 \\
		0 & \nabla^2_{\nu} f(\alpha, \hat{\beta})
	\end{bmatrix},
\end{equation}
where
\begin{equation*}
  \nabla^2_{\pi} f(\alpha, \hat{\beta}) = \text{diag} \left( \frac{\sum_{i=1}^N \hat{\tau}_{i1}}{\pi_1^2}, \ldots, \frac{\sum_{i=1}^N \hat{\tau}_{iK}}{\pi_K^2} \right)
\end{equation*}
and
\begin{equation*}
  \nabla^2_{\nu} f(\alpha, \hat{\beta}) = \text{diag} \left( -2 \hat{\eta}_m + \sum_{i=1}^N \hat{\tau}_{i1}, \ldots, -2 \hat{\eta}_m + \sum_{i=1}^N \hat{\tau}_{iK} \right).
\end{equation*}

Again, the matrix is diagonal and all of the diagonal elements are non-negative, so the Hessian is positive semidefinite and $f(\alpha, \hat{\beta})$ is convex.

Therefore, the first condition of the GOP algorithm is satisfied.
The second condition is satisfied because we have no inequality constraints that are functions of both $\alpha$ and $\beta$ -- all constraints can be absorbed into the sets $A,B$.
The third condition is satisfied because the equality constraints are each convex combinations and therefore affine.
The fourth condition (Slater's condition) is satisfied because the interior of the feasible region has an interior point.

\section{Appendix for BGMM -- Gaussian Approximation}

\subsection{Derivation of Log-likelihood Lower Bound}
\label{sec:gbgmm_elbo_derivation}

The variational lower bound is
\begin{equation}
	\begin{split}
		l_\mathcal{D} (\phi) \geq \mathcal{L}(\phi, \xi) &\triangleq E_q[\log p(y, z, m \mid \phi)] - E_q[\log q(z, m)] 
		\\
		&= \sum_{i=1}^N E_q[\log p(y_i \mid z_i, m)] + \sum_{i=1}^N E_q[\log p(z_i \mid \pi)] + \sum_{k=1}^K E_q[\log p(m_k \mid \Gamma)]
		\\
		&\quad - \sum_{i=1}^N E_q[\log q(z_i \mid \tau_i)] - \sum_{k=1}^K E_q[\log q(m_k \mid \nu_k, \gamma_k)].
	\end{split}
\end{equation}
We can write out each term in the ELBO:
\begin{align*}
	E_q[\log p(y_i \mid z_i, m)] &= E_q \left[ -\frac{1}{2} \log (2 \pi) - \frac{1}{2} y_i^2 + y_i \sum_{k=1}^K z_{ik} m_k - \frac{1}{2} \left( \sum_{k=1}^K z_{ik} m_k \right)^2 \right] \\
	&= -\frac{1}{2} \log (2 \pi) - \frac{1}{2} y_i^2 + E_q \left[ \sum_{k=1}^K z_{ik} m_k \right] y_i - \frac{1}{2} E_q \left[ \left( \sum_{k=1}^K z_{ik} m_k \right)^2 \right]
	\\
	E_q[\log p(z_i \mid \tau_i)] &= E_q \left[ \sum_{k=1}^K z_{ik} \log \tau_{ik} \right]
	=  \sum_{k=1}^K E_q [ z_{ik} ] \log \tau_{ik} 
	\\
	E_q[\log p(m_k | \eta_m)] &= E_q \left[ -\frac{1}{2} \log (2 \pi) + \frac{1}{2} \log (-2 \eta_m) + \eta_m m_k^2 \right] \\
	&= -\frac{1}{2} \log (2 \pi) + \frac{1}{2} \log (-2 \eta_m) + \eta_m E_q \left[ m_k^2 \right] 
	\\
	E_q[ \log q(z_i | \tau_i)] &= \sum_{k=1}^K \tau_{ik} \log \tau_{ik} \\
	E_q[ \log q(m_k | \nu_k, \gamma_k)] &= -\frac{1}{2} \log \left( 2 \pi e \gamma_k \right),	
\end{align*}
where we define $\eta_m = - (2 \Gamma)^{-1}$.
The ELBO requires expected values under the variational distribution for $E_q \left[ \sum_{k=1}^K z_{ik} m_k \right]$, $E_q \left[ \left( \sum_{k=1}^K z_{ik} m_k \right)^2 \right]$, $E_q[z_{ik}]$, and $E_q[m_k^2]$.
We can simplify these expected values:
\begin{align*}
	E_q \left[ \sum_{k=1}^K z_{ik} m_k \right] &= \sum_{k=1}^K E_q[z_{ik}] E_q[m_k] = \sum_{k=1}^K \tau_{ik} \nu_k \\
	E_q\left[ \left( \sum_{k=1}^K z_{ik} m_k\right)^2 \right] &= \sum_{k=1}^K E_q[z_{ik}^2] E_q[m_k^2] + 2 \sum_{k \neq k^\prime} E_q[z_{ik} z_{ik^\prime}] E_q[m_k m_{k^\prime}] \\
	&= \sum_{k=1}^K \tau_{ik} \left( \gamma_k + \nu_k^2 \right) \\
	E_q[z_{ik}] &= \tau_{ik} \\
	E_q[m_k^2] &= \gamma_k + \nu_k^2.
\end{align*}
Substituting these expectations back into the full ELBO gives us the result
\begin{multline}
	\mathcal{L}(\phi, \xi) \propto
		- \frac{1}{2} \sum_{i=1}^N \sum_{k=1}^K \tau_{ik} ( y_i - \nu_k )^2
		- \frac{1}{2} \sum_{i=1}^N \sum_{k=1}^K \tau_{ik} \gamma_k
		+ \sum_{i=1}^N \sum_{k=1}^K \tau_{ik} \log \pi_k
		+ \frac{K}{2} \log \left( -2 \eta_m \right)
		+ \eta_m \sum_{k=1}^K \nu_k^2
		+ \eta_m \sum_{k=1}^K \gamma_k
		\\
		- \sum_{i=1}^N \sum_{k=1}^K \tau_{ik} \log \tau_{ik}
		+ \frac{1}{2} \sum_{k=1}^K \log (2 \pi e \gamma_k).
\end{multline}

\subsection{Variational Inference Algorithm}\label{sec:variational_inference_algorithm}

The standard variational EM algorithm for the model is an alternating coordinate ascent algorithm on the ELBO.
The M-step maximizes the ELBO with respect to the model parameters, $\phi$.
The E-step maximizes the ELBO with respect to the variational distribution parameters, $\xi$.

\paragraph{M-step}

We can maximize the ELBO with respect to $\pi$ by setting the derivative of the ELBO function with respect to $\pi$ and solving for $\pi$.
However, we also require that $\sum_{k=1}^K \pi_k = 1$, so we add a Lagrange multiplier to enforce this constraint.
Solving for the maximum of the resulting Lagrange function gives the update formula,
\begin{equation}
	\hat{\pi}_k = \frac{1}{\lambda_p} \sum_{i=1}^N \hat{\tau}_{ik} = \frac{1}{N} \sum_{i=1}^N \hat{\tau}_{ik},
\end{equation}
We see that the MLE for $\pi$ is the usual sample mean of $\tau$.

Setting the derivative of the ELBO Lagrange function with respect to $\eta_{m}$ and solving for $\eta_{m}$ gives the update formula,
\begin{equation}
	\hat{\eta}_m = - K \left( 2 \sum_{k=1}^K \left( \nu_k^2 + \gamma_k \right) \right)^{-1}.
\end{equation}
This natural parameter estimate gives the usual MLE estimate for the corresponding moment parameter $\hat{\Gamma} = \frac{1}{K} \sum_{k=1}^K (\nu_k^2 + \gamma_k)$.

\paragraph{E-step}

Setting the derivative of the ELBO Lagrange function with respect to $\tau_{ik}$ and solving for $\tau_{ik}$ gives the update formula,
\begin{equation}
	\hat{\tau}_{ik} \triangleq \frac{1}{\lambda_{qi}} \exp \left( \nu_k - \frac{1}{2} \nu_k^2 + \eta_{z_k} \right) 
	= \frac{ \exp \left( \nu_k - \frac{1}{2} \nu_k^2 + \eta_{z_k} \right) }{ \sum_{k=1}^K \exp \left( \nu_k - \frac{1}{2} \nu_k^2 + \eta_{z_k} \right) }.
\end{equation}
where we have solved for $\lambda_{qi}$ by plugging $\hat{\tau}_{ik}$ into its constraint.
The update for $\tau_{ik}$ can be rewritten using the moment parameters as
\begin{equation*}
	\hat{\tau}_{ik} = \frac{ \pi_k \exp \left( -\frac{1}{2} (y_i - \nu_k)^2 \right) }{ \sum_{k=1}^K \pi_k \exp \left( -\frac{1}{2} (y_i - \nu_k)^2 \right) }.
\end{equation*}
This form shows that $\hat{\tau}_{ik}$ is proportional to the product of the prior for belonging to the $k$-th cluster and the likelihood of sampling $y_i$ from a Gaussian with mean $\nu_k$.

Setting the derivative of the ELBO Lagrange function with respect to $\nu_{k}$ and solving for $\nu_{k}$ gives the update formula,
\begin{equation}
	\hat{\nu}_k = \frac{ \sum_{i=1}^N y_i \hat{\tau}_{ik}  }{ \sum_{i=1}^N \hat{\tau}_{ik} - 2 \hat{\eta}_m }.
\end{equation}
The update for $\nu_k$ can be rewritten using the moment parameters as
\begin{equation*}
	\hat{\nu}_{k} = \frac{ \sum_{i=1}^N y_i \tau_{ik} }{ \sum_{i=1}^N \tau_{ik} + \Gamma^{-1} }.
\end{equation*}
This form shows that $\hat{\nu}_{k}$ is the sample mean of the weighted data assigned to the $k$-th cluster when the prior precision $\Gamma^{-1} = 0$.
This value of the precision corresponds to an improper prior on the cluster means and reduces the Bayesian Gaussian mixture model to a standard Gaussian mixture model.
When $\Gamma^{-1}$ is greater than zero $\hat{\nu}_k$ is regularized (shrunken) towards the prior mean, which in this model is zero.

Setting the derivative of the ELBO Lagrange function with respect to $\gamma_{k}$ and solving for $\gamma_{k}$ gives the update formula,
\begin{equation}
	\hat{\gamma}_k = \frac{ 1 }{ \sum_{i=1}^N \hat{\tau}_{ik} - 2 \hat{\eta}_m }.
\end{equation}

%

\subsection{Proof of GOP conditions}
\label{sec:gbgmm_biconvex_proof}

\begin{multline}
	f(\alpha, \beta) = 
		\frac{1}{2} \sum_{i=1}^N \sum_{k=1}^K \tau_{ik} ( y_i - \nu_k )^2
		+ \frac{1}{2} \sum_{i=1}^N \sum_{k=1}^K \tau_{ik} \gamma_k
		- \sum_{i=1}^N \sum_{k=1}^K \tau_{ik} \log \pi_k
		- \frac{K}{2} \log \left( -2 \eta_m \right)
		- \eta_m \sum_{k=1}^K \nu_k^2
		- \eta_m \sum_{k=1}^K \gamma_k
		\\
		+ \sum_{i=1}^N \sum_{k=1}^K \tau_{ik} \log \tau_{ik}
		- \frac{1}{2} \sum_{k=1}^K \log (2 \pi e \gamma_k),
\end{multline}
where $\alpha \triangleq \{ \pi, \nu, \gamma \}$ and $\beta \triangleq \{ \tau, \eta_m \}$.

The Hessian with respect to $\beta$ for fixed $\hat{\alpha}$ is
\begin{equation}
	\nabla_{\beta}^2 f(\hat{\alpha}, \beta) =
	\begin{bmatrix}
		\nabla^2_{\tilde{\tau}} f(\hat{\alpha}, \beta) & 0 \\
		0 & \nabla^2_{\eta_m} f(\hat{\alpha}, \beta)	
	\end{bmatrix},
\end{equation}
where 
\begin{equation*}
  \nabla^2_{\tilde{\tau}} f(\hat{\alpha}, \beta) = \text{diag} \left( \frac{1}{\tau_{11}}, \ldots, \frac{1}{\tau_{NK}} \right)
\end{equation*}
and
\begin{equation*}
  \nabla^2_{\eta_m} f(\hat{\alpha}, \beta) = \frac{K}{2 \eta_m^2}.
\end{equation*}
We define $\tilde{\tau}$ to be the matrix $\tau$ construed as a vector taken column-wise.
Since the matrix is diagonal and all of the diagonal elements are non-negative, the Hessian is positive semidefinite and $f(\hat{\alpha}, \beta)$ is convex.

The Hessian with respect to $\alpha$ for fixed $\hat{\beta}$
\begin{equation}
	\nabla_{\alpha}^2 f(\alpha, \hat{\beta}) =
	\begin{bmatrix}
		\nabla^2_{\pi} f(\alpha, \hat{\beta}) & 0 & 0\\
		0 & \nabla^2_{\nu} f(\alpha, \hat{\beta}) & 0 \\
		0 & 0 & \nabla^2_{\gamma} f(\alpha, \hat{\beta})
	\end{bmatrix},
\end{equation}
where
\begin{equation*}
  \nabla^2_{\pi} f(\alpha, \hat{\beta}) = \text{diag} \left( \frac{\sum_{i=1}^N \hat{\tau}_{i1}}{\pi_1^2}, \ldots, \frac{\sum_{i=1}^N \hat{\tau}_{iK}}{\pi_K^2} \right),
\end{equation*}
\begin{equation*}
  \nabla^2_{\nu} f(\alpha, \hat{\beta}) = \text{diag} \left( -2 \hat{\eta}_m + \sum_{i=1}^N \hat{\tau}_{i1}, \ldots, -2 \hat{\eta}_m +\sum_{i=1}^N \hat{\tau}_{iK} \right),
\end{equation*}
and
\begin{equation*}
  \nabla^2_{\gamma} f(\alpha, \hat{\beta}) = \text{diag} \left( \frac{1}{2 \gamma_1^2}, \ldots, \frac{1}{2 \gamma_K^2} \right).
\end{equation*}

Again, the matrix is diagonal and all of the diagonal elements are non-negative, so the Hessian is positive semidefinite and $f(\alpha, \hat{\beta})$ is convex.

Therefore, the first condition of the GOP algorithm is satisfied.
The second condition is satisfied because we have no inequality constraints that are functions of both $\alpha$ and $\beta$ -- all constraints can be absorbed into the sets $A,B$.
The third condition is satisfied because the equality constraints are each convex combinations and therefore affine.
The fourth condition (Slater's condition) is satisfied because the interior of the feasible region has an interior point.

\newpage
\small
\bibliographystyle{abbrvnat}
\bibliography{gop-bib}

\begin{thebibliography}{15}
\providecommand{\natexlab}[1]{#1}
\providecommand{\url}[1]{\texttt{#1}}
\expandafter\ifx\csname urlstyle\endcsname\relax
  \providecommand{\doi}[1]{doi: #1}\else
  \providecommand{\doi}{doi: \begingroup \urlstyle{rm}\Url}\fi

\bibitem[Archambeau(2003)]{archambeau2003}
V.~M. Archambeau, Lee A~J.
\newblock On convergence problems of the {EM} algorithm for finite {G}aussian
  mixtures.
\newblock \emph{European Symposium on Artificial Neural Networks}, pages
  99--106, 2003.

\bibitem[Beal and Ghahramani(2003)]{Beal2003a}
M.~J. Beal and Z.~Ghahramani.
\newblock The variational {B}ayesian {EM} algorithm for incomplete data: with
  application to scoring graphical model structures.
\newblock \emph{Bayesian Statistics}, 7, 2003.

\bibitem[Benders(1962)]{Benders1962a}
J.~F. Benders.
\newblock Partitioning procedures for solving mixed-variables programming
  problems.
\newblock \emph{Computational Management Science}, 2\penalty0 (1):\penalty0
  3--19, Jan. 1962.

\bibitem[Dempster et~al.(1977)Dempster, Laird, and Rubin]{Dempster1977a}
A.~P. Dempster, N.~M. Laird, and D.~B. Rubin.
\newblock Maximum likelihood from incomplete data via the em algorithm.
\newblock \emph{Journal of the Royal Statistical Society. Series B
  (Methodological)}, 39\penalty0 (1):\penalty0 1--38, 1977.
\newblock ISSN 00359246.

\bibitem[Floudas(2000)]{Floudas2000a}
C.~A. Floudas.
\newblock \emph{Deterministic Global Optimization}.
\newblock Theory, Methods and Applications. Springer US, 2000.

\bibitem[Floudas and Visweswaran(1990)]{Floudas1990a}
C.~A. Floudas and V.~Visweswaran.
\newblock A global optimization algorithm ({GOP}) for certain classes of
  nonconvex {NLP}s{\textemdash}{I}. theory.
\newblock \emph{Computers {\&} chemical engineering}, 14\penalty0
  (12):\penalty0 1397--1417, Dec. 1990.

\bibitem[Floudas and Visweswaran(1993)]{Floudas1993a}
C.~A. Floudas and V.~Visweswaran.
\newblock Primal-relaxed dual global optimization approach.
\newblock \emph{Journal of Optimization Theory and Applications}, 78\penalty0
  (2):\penalty0 187--225, 1993.

\bibitem[Gelman et~al.(2013)Gelman, Carlin, Hal, David, Vehtari, and
  Donald]{Gelman2013a}
A.~Gelman, J.~B. Carlin, S.~Hal, D.~David, A.~Vehtari, and R.~Donald.
\newblock \emph{Bayesian Data Analysis}.
\newblock Chapman and Hall, 3 edition, 2013.

\bibitem[Geoffrion(1972)]{Geoffrion1972a}
A.~M. Geoffrion.
\newblock {Generalized {B}enders decomposition}.
\newblock \emph{Journal of optimization theory and applications}, 1972.

\bibitem[Hansen et~al.(1993)Hansen, Jaumard, and Xiong]{Hansen1993a}
P.~Hansen, B.~Jaumard, and J.~Xiong.
\newblock Decomposition and interval arithmetic applied to global minimization
  of polynomial and rational functions.
\newblock \emph{Journal of Computational Optimization}, 3\penalty0
  (4):\penalty0 421--437, Dec 1993.

\bibitem[Murray(1977)]{Murray1977a}
G.~D. Murray.
\newblock Contribution to discussion of paper by {A. P. Dempstar, N. M. Laird,
  and D. B. Rubin}.
\newblock \emph{Journal of the Royal Statistical Society. Series B
  (Methodological)}, 39:\penalty0 27--28, 1977.

\bibitem[Sahinidis(1996)]{sahinidis1996baron}
N.~V. Sahinidis.
\newblock {BARON}: A general purpose global optimization software package.
\newblock \emph{Journal of global optimization}, 8\penalty0 (2):\penalty0
  201--205, 1996.

\bibitem[Tawarmalani and Sahinidis(2005)]{Tawarmalani2005a}
M.~Tawarmalani and N.~V. Sahinidis.
\newblock A polyhedral banch-and-cut approach to global optimization.
\newblock \emph{Mathematical Programming}, 103\penalty0 (2):\penalty0 225--249,
  2005.

\bibitem[Visweswaran and Floudas(1996)]{Visweswaran1996-mr}
V.~Visweswaran and C.~A. Floudas.
\newblock New formulations and branching strategies for the {GOP} algorithm.
\newblock In I.~E. Grossmann, editor, \emph{Global Optimization in Engineering
  Design}, Nonconvex Optimization and Its Applications, pages 75--109. Springer
  US, 1996.

\bibitem[Wu(1983)]{wu1983}
C.~F. Wu.
\newblock On the convergence properties of the {EM}.
\newblock \emph{The Annals of Statistics}, 11\penalty0 (1):\penalty0 95--103,
  1983.

\end{thebibliography}
\end{document}